\documentclass[prc,showpacs,showkeys,superscriptaddress,nofootinbib,twocolumn,floatfix]{revtex4}
\usepackage{amsfonts}
\usepackage{graphicx,color,amsmath,amssymb}

\def\blfootnote{\xdef\@thefnmark{}\@footnotetext}

\newcommand{\gsim}{\gtrsim}

\newcommand{\vecp}{\mathbf{p}}
\newcommand{\vecx}{\mathbf{x}}

\begin{document}

\title{Heavy-Quark Diffusion and Hadronization in Quark-Gluon Plasma}

\author{Min He}
\affiliation{Cyclotron Institute and Department of Physics and Astronomy, Texas A\&M University, College Station, TX 77843, USA}

\author{Rainer J.\ Fries}
\affiliation{Cyclotron Institute and Department of Physics and Astronomy,
  Texas A\&M University, College Station, TX 77843, USA}
\affiliation{RIKEN/BNL Research Center, Brookhaven National Laboratory,
Upton, NY 11973, USA}

\author{Ralf Rapp}
\affiliation{Cyclotron Institute and Department of Physics and Astronomy, Texas A\&M University, College Station, TX 77843, USA}

\date{\today}

\begin{abstract}
We calculate diffusion and hadronization of heavy quarks in
high-energy heavy-ion collisions implementing the notion of a
strongly coupled quark-gluon plasma in both micro- and macroscopic
components. The diffusion process is simulated using relativistic
Fokker-Planck dynamics for elastic scattering in a hydrodynamic
background. The heavy-quark transport coefficients in the medium are
obtained from non-perturbative $T$-matrix interactions which build
up resonant correlations close to the transition temperature. The
latter also form the basis for hadronization of heavy quarks into
heavy-flavor mesons via recombination with light quarks from the
medium. The pertinent resonance recombination satisfies energy
conservation and provides an equilibrium mapping between quark and
meson distributions. The recombination probability is derived from
the resonant heavy-quark scattering rate. Consequently,
recombination dominates at low transverse momentum ($p_T$) and
yields to fragmentation at high $p_T$. Our approach thus emphasizes
the role of resonance correlations in the diffusion and
hadronization processes. We calculate the nuclear modification
factor and elliptic flow of $D$- and $B$-mesons for Au-Au collisions
at the Relativistic Heavy Ion Collider, and compare their
decay-electron spectra to available data. We also find that a
realistic description of the medium flow is essential for a
quantitative interpretation of the data.
\end{abstract}

\maketitle

\section{Introduction}
\label{sec_intro}
Experiments at the Relativistic Heavy Ion Collider (RHIC) and the
Large Hadron Collider (LHC) have been searching for the deconfined
phase of nuclear matter and have begun to probe its
properties~\cite{Adams:2005dq,Schukraft:2011kc}. There are strong
indications that this new form of matter behaves like a nearly
perfect fluid with high opacity and low viscosity, referred to as
strongly coupled Quark-Gluon Plasma
(sQGP)~\cite{Gyulassy:2004zy,Shuryak:2008eq}. One of the major
experimental findings is a large azimuthal anisotropy, $v_2$, in
transverse momentum ($p_T$) spectra of hadrons in non-central
collisions~\cite{Voloshin:2008dg}. To account for this observation,
hydrodynamic simulations require an early initialization time,
implying a rapid thermalization of the bulk
medium~\cite{Teaney:2001av,Heinz:2009xj,Hirano:2010je}.
However, the microscopic origin of the rapid thermalization remains a
matter of debate.

In contrast to light partons making up the bulk of the medium, heavy
quarks (charm and bottom), produced in primordial hard collisions
and acting as impurities in the QGP, are not expected to fully
equilibrate with the surrounding medium. Due to their large masses
($m_Q$) a memory of their interaction history may be preserved, thus
providing a more direct probe of the medium properties than bulk
observables~\cite{Svetitsky:1987gq,vanHees:2004gq,Moore:2004tg,Rapp:2009my}.
The thermal relaxation time of heavy quarks has been argued to be
larger than that of light quarks by a factor of $m_Q/T\approx
5-20$~\cite{Moore:2004tg,Rapp:2009my} ($T$: typical temperature of
the QGP). As they diffuse through the medium, heavy quarks interact
with the light partons and their spectrum becomes
quenched~\cite{vanHees:2004gq,Moore:2004tg}. Moreover, as they
couple to the collective flow of the medium in non-central heavy-ion
collisions, heavy quarks may develop substantial momentum
anisotropies. These two effects are translated into equivalent
behavior of heavy-flavor (HF) meson ($D$ and $B$) spectra and $v_2$,
and further into the spectrum and $v_2$ of their decay electrons.
The latter have been measured in Au+Au collisions at
RHIC~\cite{Abelev:2006db,Adare:2006nq,Adare:2010de}, exhibiting
appreciable modifications over their baseline spectra from $p+p$ and
$d+$Au collisions.

Model calculations based on radiative energy loss in perturbative QCD
(pQCD), which could account for the observed jet-quenching in the light
sector~\cite{Gyulassy:2003mc}, predicted a much smaller quenching for
heavy quarks and associated single-electron spectra~\cite{Wicks:2005gt}.
The large HQ mass suppresses small-angle gluon radiation (``dead cone"
effect~\cite{Dokshitzer:2001zm}) and reduces the gluon formation
time~\cite{Zhang:2003wk}, hence mitigating radiative energy loss
significantly. However, elastic collisions of heavy quarks with light
partons~\cite{Svetitsky:1987gq,Braaten:1991we,vanHees:2004gq,Moore:2004tg,Mustafa:2004dr}
have been argued to dominate over radiative scattering at low momentum,
resulting in notable quenching of the HQ spectrum.

However, jet quenching only captures part of the physics potential
of the HQ probe. Its diffusion properties, which reach all the way
to zero momentum, include energy-gain processes which are, e.g.,
instrumental for the coupling to the collective flow of the medium.
Several studies of HQ diffusion have been conducted in recent years
using
Fokker-Planck~\cite{vanHees:2004gq,Moore:2004tg,Mustafa:2004dr,vanHees:2005wb,vanHees:2007me,Gossiaux:2008jv,Akamatsu:2008ge,Das:2009vy,Alberico:2011zy}
and Boltzmann
transport~\cite{Zhang:2005ni,Molnar:2006ci,Uphoff:2010sh}
approaches, mostly implementing elastic collisions as the
microscopic dynamics. They differ not only in their treatment of the
background medium, but also in the evaluation of (a) the transport
coefficients emerging from the interactions between the heavy quarks
and the medium, and (b) hadronization of heavy quarks into HF
mesons. Concerning item (a), most studies employ variants of the
pQCD interaction~\cite{Combridge:1978kx}, while a novel approach
with heavy-light resonant interactions was introduced in
Refs.~\cite{vanHees:2004gq,vanHees:2005wb}. The latter was found to
be a factor of 3-4 more efficient in HQ thermalization than pQCD,
and was subsequently corroborated by microscopic $T$-matrix
calculations using input potentials from lattice QCD
(lQCD)~\cite{vanHees:2007me,Riek:2010fk,Riek:2010py}. Concerning HQ
hadronization, several studies focused on independent
fragmentation~\cite{Akamatsu:2008ge,Das:2009vy,Alberico:2011zy,Uphoff:2010sh},
which is not reliable in the low and intermediate-$p_T$ regimes.
Here, light partons surrounding the heavy quark have a high
phase-space density which renders coalescence a more plausible
hadronization
mechanism~\cite{Lin:2003jy,Fries:2003vb,Greco:2003mm,Fries:2008hs}.
In Refs.~\cite{vanHees:2005wb,vanHees:2007me}, heavy-light quark
recombination has been incorporated utilizing an instantaneous
coalescence model~\cite{Greco:2003vf} which could still be
problematic at low $p_T$ due to lack of energy conservation. A
reliable treatment of the low-$p_T$ regime is important since the
total number of heavy quarks is expected to be conserved through the
hadronization transition. If the $D$- or $B$-meson spectra are
distorted at low $p_T$, the spectra at higher $p_T$ are necessarily
affected thus modifying the $R_{AA}$ (and $v_2$) of $D$ and
$B$-mesons and their decay electrons.

The purpose of the present work is to establish a realistic and
quantitative framework for HQ probes within (a) a strongly coupled
QGP background medium (modeled by hydrodynamics), (b) a
non-perturbative scenario of elastic diffusion in the QGP simulated
by Fokker-Planck-Langevin dynamics, and (c) a hadronization scheme
at the phase transition based on the same interaction as in (b),
combining recombination and fragmentation consistent with the
limiting cases of kinetic equilibrium and vacuum hadronization.
Unlike previous studies utilizing weak-coupling
diffusion~\cite{Moore:2004tg,Gossiaux:2008jv,Das:2009vy,Alberico:2011zy,Uphoff:2010sh}
we try to implement the HQ probe consistently within a framework of
strong coupling between heavy and light quarks, both in the QGP and
during hadronization. Our comprehensive framework is hence
conceptually compatible with the notion of a strongly interacting
QGP.

The strategy in this work is as follows. For the HQ transport
coefficient we employ a non-perturbative $T$-matrix calculation of
heavy-light quark interactions~\cite{Riek:2010fk,Riek:2010py}.
This calculation supports Feshbach resonances in the QGP in the
color-singlet and anti-triplet channels, surviving as rather broad
states up to $\sim1.5~T_c$. They are responsible for the enhancement
of the transport coefficient compared to pQCD scattering. With these
coefficients we perform Langevin simulations of HQ diffusion through
an expanding medium which is described by ideal 2+1-dimensional
hydrodynamics (using the AZHYDRO code~\cite{Kolb:2003dz} at RHIC
energies). At the phase transition, heavy quarks are hadronized through
coalescence with light quarks of the medium using the Resonance
Recombination Model (RRM)~\cite{Ravagli:2007xx} implemented on a
hypersurface given by the hydrodynamic simulation. The coalescence
probability is evaluated using the resonant scattering rate of the
heavy quark with light (anti) quarks, supplemented by independent
fragmentation. The RRM formalism is consistent with the heavy-light
Feshbach resonance formation found in the $T$-matrix used for the
transport coefficient. This stipulates the role played by the resonance
correlations in our work. With an artificially large transport
coefficient, we check the equilibrium limit of the HQ distribution
emanating from the combined hydro+Langevin simulation and the ensuing
degree of equilibration of the HF mesons upon resonance recombination.
The full space-momentum correlations generated by the hydro-Langevin
simulation enter into resonance recombination. This enables a
quantitative assessment of the radial medium flow on HF meson spectra
at low $p_T$ as imprinted on the final $R_{AA}$ measurement.

Our article is organized as follows. In Sec.~\ref{sec_Langevin} we
introduce the ingredients for the hydro-Langevin simulation of HQ
diffusion in the medium, i.e., the transport coefficient, the initial
distribution in coordinate and momentum space, and the background medium
described by an ideal hydrodynamic model.  Numerical results for the HQ
$R_{AA}$ and $v_2$ are discussed in the equilibrium limit as well as for
realistic coefficients.  Sec.~\ref{sec_hadronization} is devoted to HQ
hadronization. We implement the RRM formalism on arbitrary hadronization
hypersurfaces, elaborate the equilibrium mapping in resonance
recombination, and determine the partition of coalescence and
fragmentation.  In Sec.~\ref{sec_flow} we examine consequences of
modifying the medium flow for the predicted HF meson spectra, triggered
by indications that the partonic flow of the hydrodynamic evolution is
too soft.  In Sec.~\ref{sec_electron} we make contact with current
experiments in terms of the nuclear modification factor and the elliptic
flow of electrons from HF decays. In Sec.~\ref{summary} we summarize and
conclude.

\section{Langevin Simulation of Heavy Quark Diffusion}
\label{sec_Langevin}

\subsection{Relativistic Langevin Kinetics}
\label{ssec_2.1}
The thermal momentum of a heavy quark at temperatures characteristic
for heavy-ion collisions at RHIC amounts to $p_{th}\sim
\sqrt{m_QT}$, which is parametrically larger than the typical
momentum transfer, $q\sim T$, in a single elastic collision with a
light parton from the bulk medium. Therefore many collisions are
needed to change the HQ momentum
considerably~\cite{vanHees:2004gq,Moore:2004tg}. This forms the
basis for approximating the HQ motion in the QGP by a succession of
uncorrelated momentum kicks and leads to a Fokker-Planck approach
realized stochastically by the Langevin
equations~\cite{Landau1981,Svetitsky:1987gq,Rapp:2009my,Hanggi2009}
\begin{align}
\label{Langevinrule1}
d\vecx&=\frac{\vecp}{E}dt,\\
d\vecp&=-\Gamma(p) \ \vecp \ dt + \sqrt{2D(\vecp+d \vecp) \, dt} \
\mathbf{\rho} \ , \label{Langevinrule2}
\end{align}
where $\vecx$ and $\vecp$ are the position and momentum vector of
the heavy quark, and $E(p)=(m_Q^2+\vecp^2)^{1/2}$ is its energy. In
the following we employ the post-point discretization scheme in
which the equilibrium condition (the relativistic
fluctuation-dissipation theorem) takes the simple form
\begin{equation}
D(p)=\Gamma(p)~E(p)~T
\label{equilcond}
\end{equation}
with $\Gamma(p)$ being the drag coefficient and $D(p)$ the (diagonal)
diffusion coefficient. The standard Gaussian noise variable,
$\mathbf{\rho}$, is distributed according to
\begin{equation}
  w(\mathbf{\rho})=\frac{1}{(2\pi)^{3/2}}e^{-\mathbf{\rho}^2/2} \, .
\end{equation}
Neither the original Fokker-Planck
equation~\cite{Landau1981,Rapp:2009my} nor the Langevin equation is
Lorentz covariant. We choose the momentum and position updates for our
HQ test particles to be at equidistant time steps $d\tau$ in the lab
frame. For a flowing medium, as in our context, the momentum updates
are rather to be done in the fluid rest frame. The updated 4-momentum
is boosted back to the lab frame with the fluid four-velocity
$u^{\mu}(x)=\gamma(v)(1,\mathbf{v}(x))$. The aforementioned equilibrium
condition must be satisfied in order for the long-time limit of the
test particle distribution to converge to the equilibrium
(Boltzmann-J\"uttner) distribution as defined by the underlying
background medium. Further details of our algorithm will be
detailed in a forthcoming article~\cite{langevin}.

\subsection{Thermal Relaxation Rate of Heavy Quarks}
\label{ssec_2.2}
The transport coefficient most commonly calculated from an underlying
microscopic interaction of the heavy quark with the bulk medium is
the thermal relaxation rate $A(p;T)$. It is related to the drag
coefficient, $\Gamma(p;T)$, in the post-point Langevin scheme,
Eq.~(\ref{Langevinrule2}), through
\begin{equation}
  \Gamma(p;T)=A(p;T)+\frac{1}{E(p)}\frac{\partial
  D(p;T)}{\partial E(p)} \ .
\label{Gamma-A}
\end{equation}
Utilizing the equilibrium condition (\ref{equilcond}) one can argue
that $\Gamma(p) = A(p) + \mathcal{O}(T/m_Q)$ and neglect terms
to higher order in the inverse HQ mass (relative to the medium
temperature).

\begin{figure}[!t]
\includegraphics[width=\columnwidth]{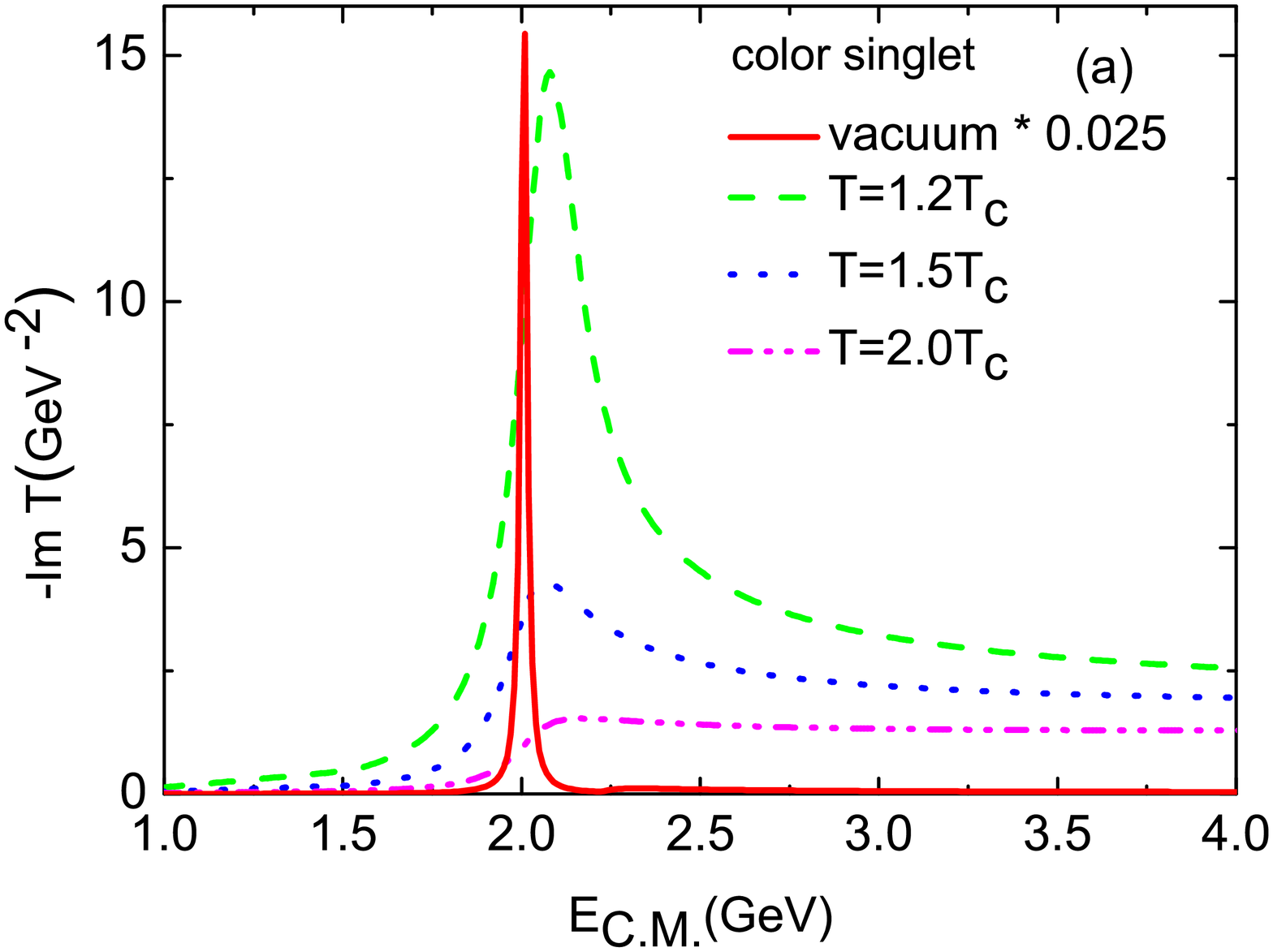}
\includegraphics[width=\columnwidth]{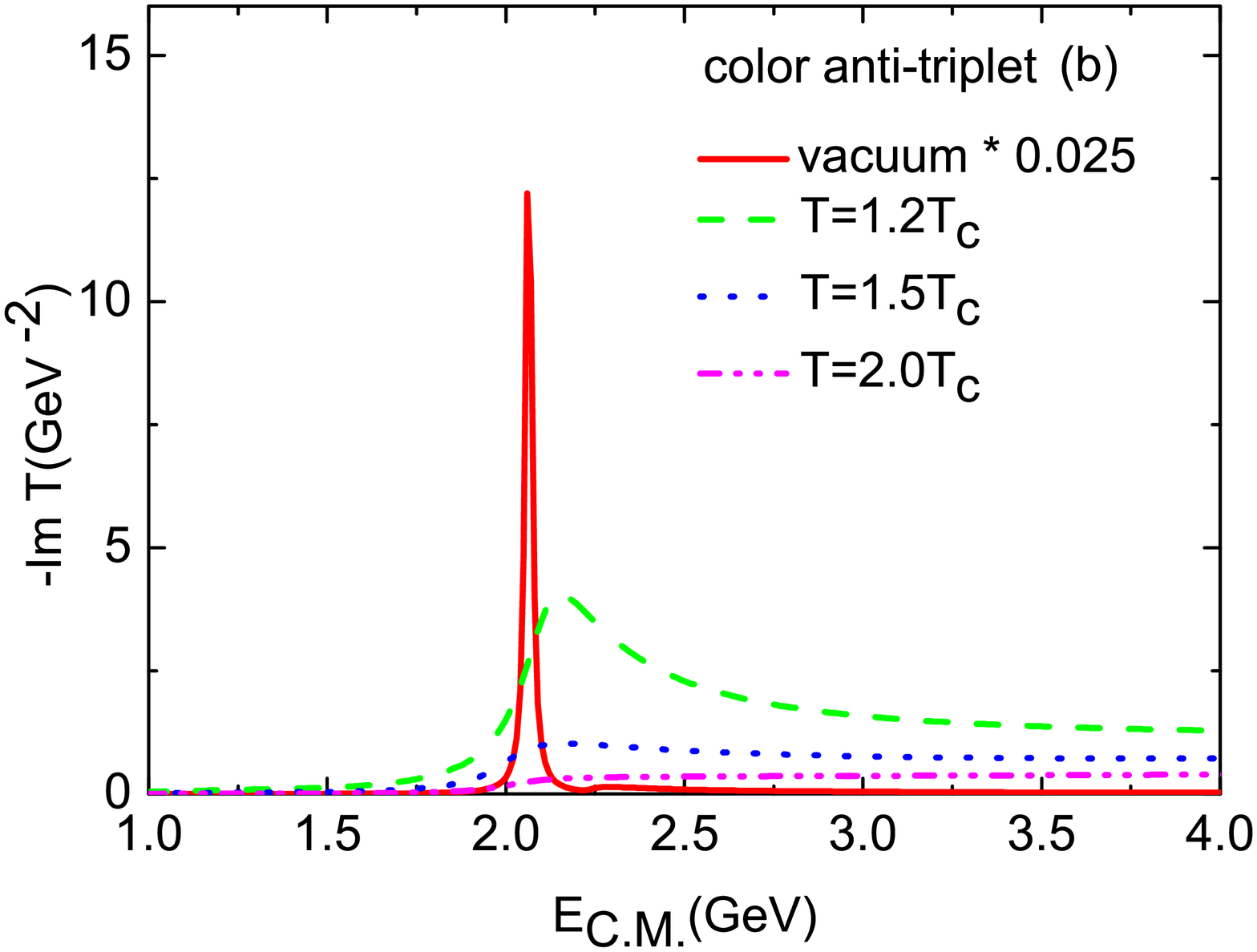}
\caption{(Color online) Imaginary part of the in-medium on-shell
$T$-Matrix for charm-light quark scattering as a function of
center-of-mass energy in the color-singlet (upper panel) and
anti-triplet (lower panel) channels, taken from the lattice-QCD based
potential approach of Ref.~\cite{Riek:2010fk}. The vacuum $T$-matrices
have been downscaled by a factor of $0.025$.}
\label{fig_Tmat}
\end{figure}
\begin{figure}[!t]
\hspace{4mm}
\includegraphics[width=\columnwidth
]{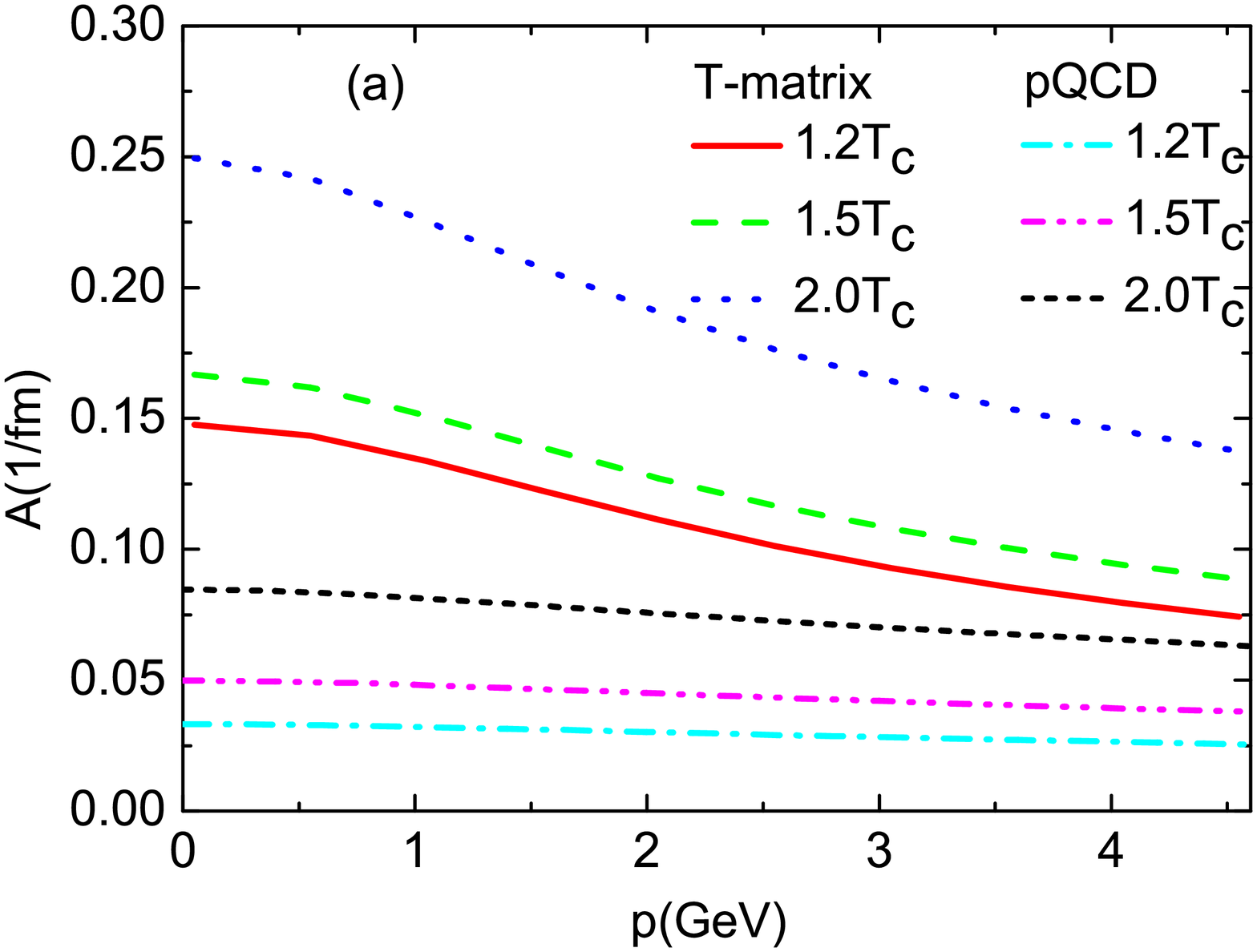}
\includegraphics[width=\columnwidth
]{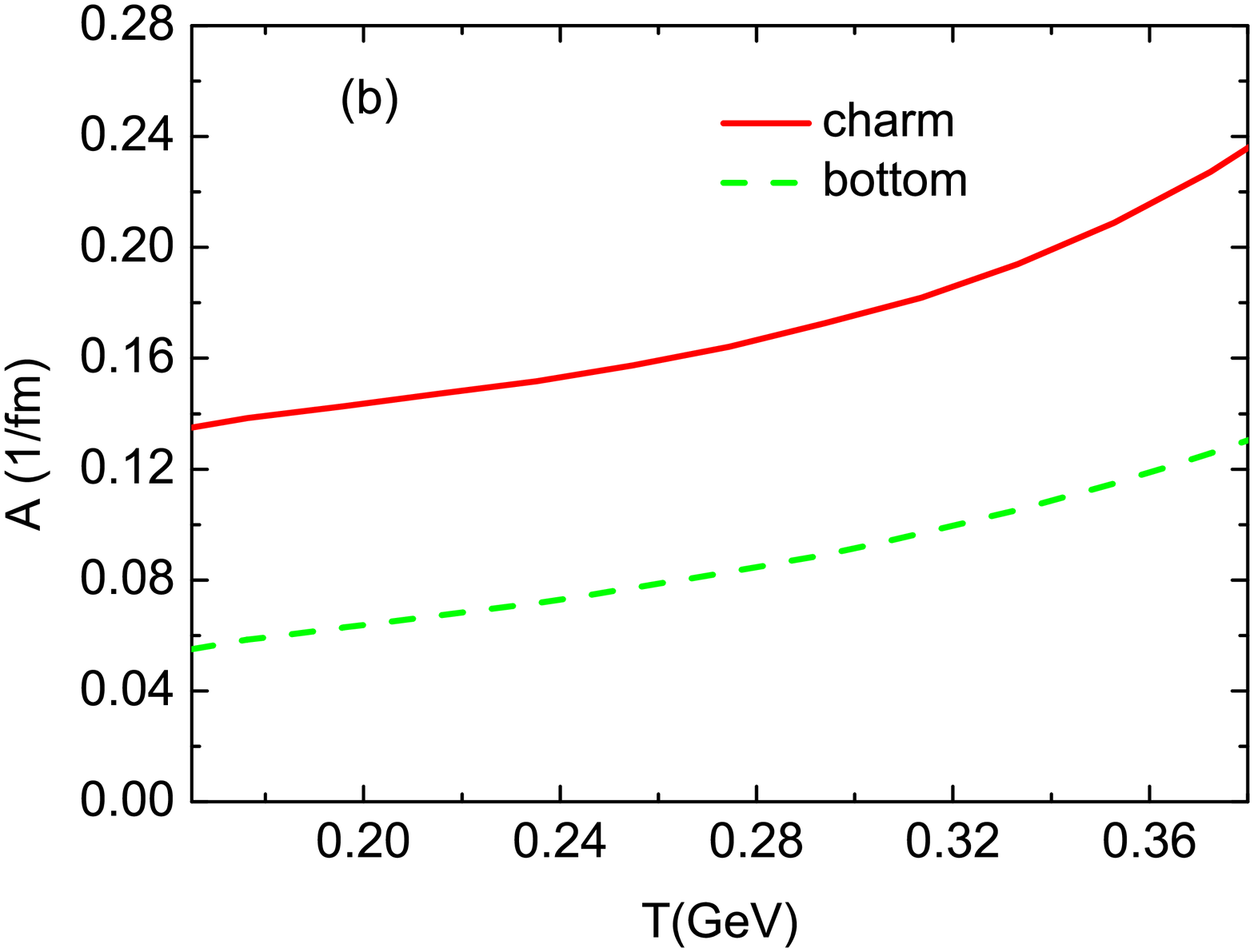} \caption{(Color online) (a)
Charm-quark relaxation rate as a function of three-momentum using
(i) heavy-light quark $T$-matrices (with lQCD internal
energy~\cite{Kaczmarek:2005ui} as potential) plus pQCD gluon
scattering with $\alpha_s=0.4$ (upper 3 curves), and (ii) pQCD
scattering off anti-/quarks and gluons with $\alpha_s=0.4$ (lower 3
curves). (b) Temperature dependence of the charm/bottom quark
thermal relaxation rate (at vanishing momentum) used in our
simulations. The results are taken from Ref.~\cite{Riek:2010fk}.}
\label{fig_Ap}
\end{figure}
We employ HQ relaxation rates from Refs.~\cite{Riek:2010fk}, where
in-medium $T$-matrices have been calculated for both heavy-light and
quarkonium channels. The input potentials were constructed using a
field-theoretic ansatz for a confining and a color-Coulomb
interaction with parameters fitted to color-average free energies
computed in finite-temperature lattice QCD
(lQCD)~\cite{Kaczmarek:2005ui}. This approach treats heavy quarkonia
and heavy-light interactions in the QGP on an equal footing, and in
both bound-state and scattering regimes. One thus obtains mutual
constraints by analyzing, e.g., Euclidean correlation functions and
HQ susceptibilities which turn out to agree fairly well with thermal
lQCD ``data"~\cite{Riek:2010py}. For heavy-light quark scattering,
the (non-perturbative) resummation in the $T$-matrix generates
resonances close to the 2-particle threshold (commonly referred to
as ``Feshbach resonances") in the attractive color-singlet (meson)
and color-anti-triplet (diquark) channels up to temperatures of
about 1.5\,$T_c$, see Fig.~\ref{fig_Tmat} for charm quarks (similar
results are obtained in the bottom sector). The increasing strength
of the $T$-matrices in the color-singlet and anti-triplet channels
when approaching $T_c$ from above is indicative for ``pre-hadronic"
correlations leading to hadronization. But even at high temperatures
a substantial enhancement of the $T$-matrix over elastic pQCD
amplitudes persists, in particular close to threshold. The rather
large resonance widths are mostly generated through the
self-energies of the light- and heavy-quark propagators in the
$T$-matrix (evaluated self-consistently in the HQ sector).

The $T$-matrices have been used to calculate thermal relaxation rates of
heavy quarks~\cite{Riek:2010fk,Riek:2010py}. Resonant rescattering
accelerates kinetic equilibration by up to a factor of
$\sim$3-5 relative to leading order (LO) pQCD
calculations~\cite{Svetitsky:1987gq}, cf.~upper panel of
Fig.~\ref{fig_Ap}. With increasing HQ 3-momentum the thermal phase space
of comoving partons (suitable for forming a Feshbach resonance)
decreases and the relaxation rate approaches the pQCD results.
For high energies and in Born approximation the $T$-matrix results
recover the LO pQCD scattering amplitudes~\cite{Riek:2010fk}. The
temperature dependence of the charm and bottom relaxation
rates (at vanishing 3-momentum) used in our simulations is
displayed in the lower panel of Fig.~\ref{fig_Ap}. They have been
extrapolated linearly from the transition temperature in the lQCD
calculations of the free energies~\cite{Kaczmarek:2005ui},
$T_c$=196\,MeV, to $T_c$=165\,MeV implicit in the equation of
state as used in AZHYDRO.

\subsection{The hydrodynamic background QGP medium}
\label{ssec_2.3}
Hydrodynamic simulations are widely applied to model the bulk
evolution of the matter created in heavy-ion collisions at
RHIC~\cite{Teaney:2001av,Heinz:2009xj,Hirano:2010je}, providing a
good description of hadron spectra and their elliptic flow. Here we
use a hydrodynamic simulation of the fireball to provide the
background medium for HQ diffusion. It supplies the information on
the space-time evolution of energy and entropy density, as well as
temperature and fluid velocity which are needed to calculate the
transport coefficients in the Langevin dynamics and on the
hadronization hypersurface. We have employed the publicly available
ideal 2+1-dimensional AZHYDRO code~\cite{Kolb:2003dz} in our study.
It assumes longitudinal boost invariance~\cite{Bjorken:1982qr} and
has been tuned to fit to bulk observables at kinetic freeze-out at
an energy density of $e_{\rm fo}=0.075~{\rm GeV/fm^3}$ in
$\sqrt{s_{NN}}=200~{\rm GeV}$ Au+Au collisions at
RHIC~\cite{Kolb:2003dz}. The initialization of AZHYDRO is done at
$\tau_0=0.6~{\rm fm/c}$ by specifying the entropy density
distribution as
\begin{equation}
\label{AZHYDROentropy}
s(\tau_0,x,y;b)=\kappa[\frac{1}{4}n_{\rm BC}(x,y;b)
+ \frac{3}{4} n_{\rm WN}(x,y;b)] \ ,
\end{equation}
where $n_{\rm BC}$ and $n_{\rm WN}$ are the binary-collision and
wounded-nucleon densities, respectively, calculated in the optical
Glauber model~\cite{Kolb:2003dz}, and $b$ is the impact parameter. The
coefficient $\kappa$ is fitted to the observed rapidity
density of charged hadrons, $dN_{\rm ch}/dy$, and translates into an
initial entropy density of $s(\tau_0,0,0;0)=110{\rm /fm^3}$ at
the center of the transverse plane for central Au+Au collisions at RHIC.

\begin{figure}[!t]
\hspace{4mm}
\includegraphics[width=\columnwidth
]{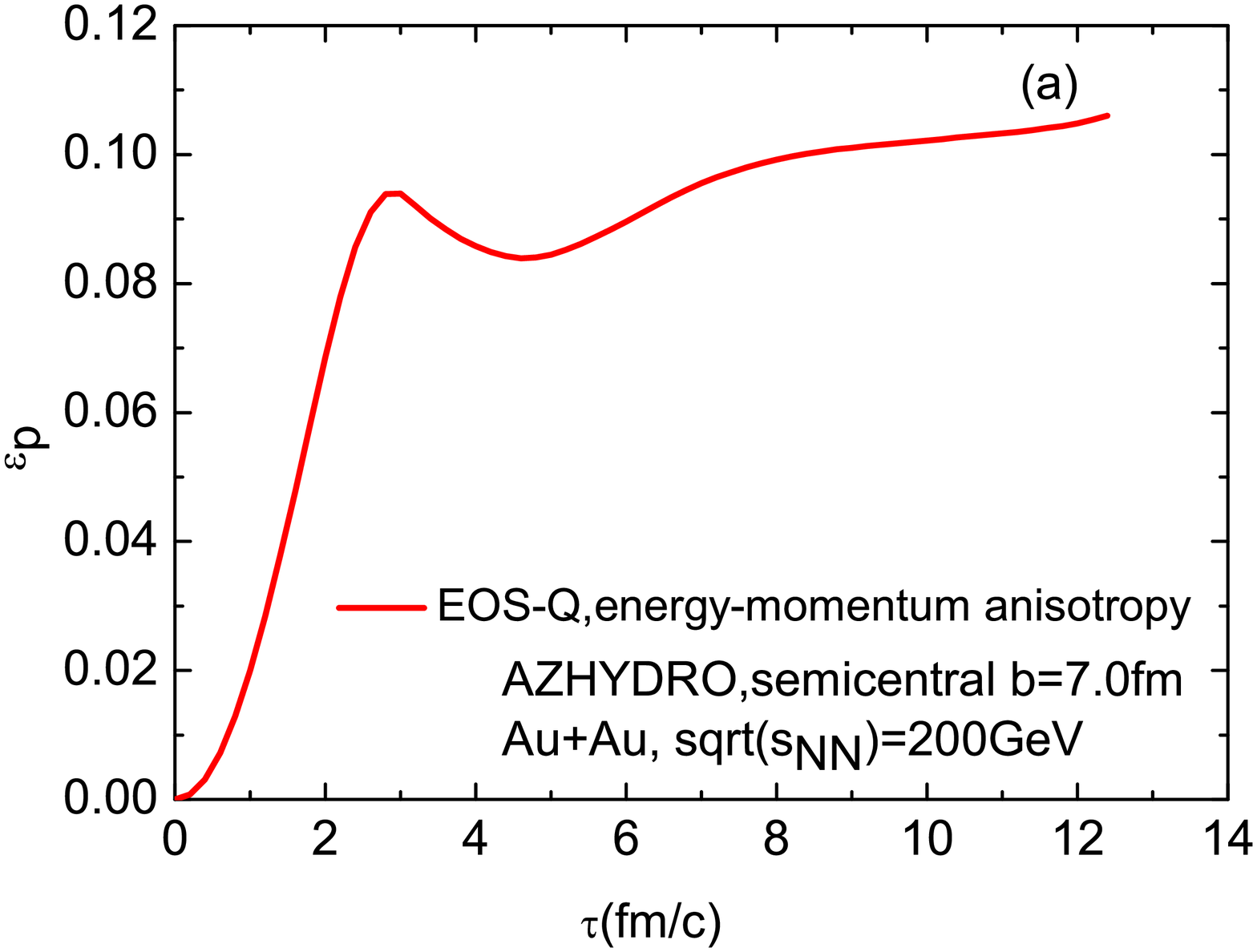}
\includegraphics[width=\columnwidth
]{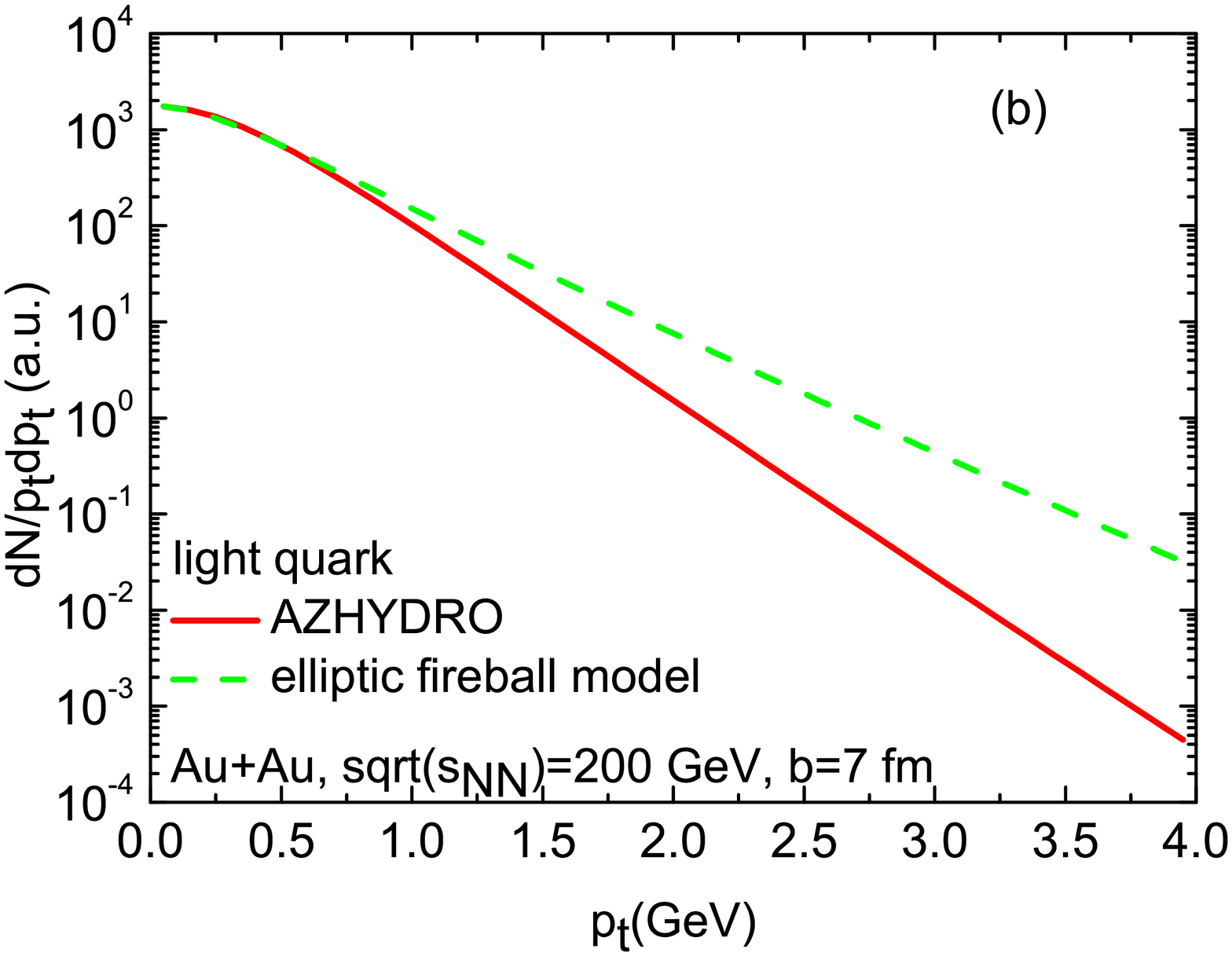}
\includegraphics[width=\columnwidth
]{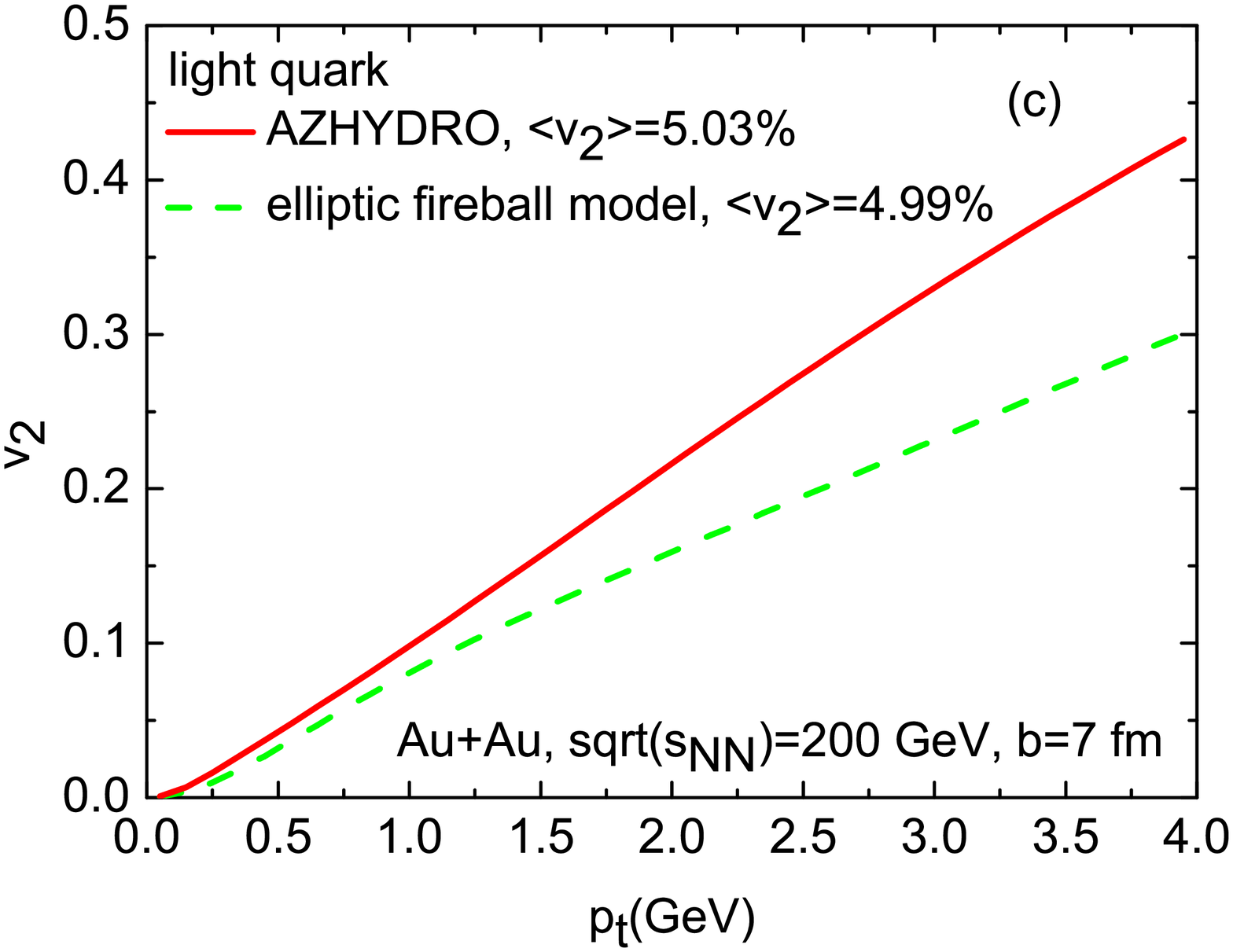} \caption{(Color online) (a) The time
evolution of the asymmetry $\epsilon_p$ of the energy-momentum
tensor in AZHYDRO for $b=7$ fm Au+Au collisions at $\sqrt{s_{\rm NN}} = 200$ GeV.
(b) Light quark ($m_q=350~{\rm MeV}$)
$p_t$-spectrum calculated with freeze-out at the end of the mixed
phase with decoupling energy density $e_{\rm dec}=0.445~{\rm
GeV/fm^3}$ in AZHYDRO (red solid line). It is compared to the light
quark spectrum at the end of the mixed phase of the parameterized
elliptic fireball model discussed in Sec.~\ref{sec_flow} (green
dashed line). (c) The light quark elliptic flow $v_2$ at the end of the
mixed phase. Again, AZHYDRO and the parameterized fireball results are
shown.} \label{fig_AZHYDRO}
\end{figure}

In Fig.~\ref{fig_AZHYDRO} we summarize the main features of AZHYDRO
relevant to our HQ diffusion calculations. The upper panel displays
the time evolution of the energy-momentum anisotropy,
$\epsilon_p=\langle T^{xx}-T^{yy}\rangle/\langle
T^{xx}+T^{yy}\rangle$ for semi-central collisions ($b$=7\,fm); it
exhibits the development of the bulk anisotropy which leads to an
elliptic flow for final-state particles~\cite{Kolb:2003dz}. One sees
that $\epsilon_p$ tends to saturate at later times when the spatial
anisotropy of the system has essentially vanished; the dip around
$\tau\simeq5$~fm/$c$ is due to the vanishing acceleration in the
mixed phase, which, in turn, is a result of the equation of state
(EoS) with a Maxwell construction between a non-interacting QGP with
a bag constant, $B=0.3642~{\rm GeV/fm^3}$ at $T>T_c=165$\,MeV, and a
hadronic resonance gas at $T<T_c$. Since in our Langevin simulations
the HQ test particles freeze out at the end of the mixed phase (at
$e_{\rm dec}=0.445~{\rm GeV/fm^3}$), we show in the middle and lower
panels of Fig.~\ref{fig_AZHYDRO} the light-quark $p_t$-spectrum and
$v_2$ at this point, respectively. The light-quark mass is taken as
$m_q=350$ MeV and we used the standard Cooper-Frye freeze-out
procedure~\cite{Cooper:1974mv,Kolb:2003dz}. For comparison we also
show the results of an empirical fireball parametrization of quark
distributions extracted from multi-strange hadron spectra in
Ref.~\cite{He:2010vw}. The quark-$p_t$ spectra are noticeably harder
than in the hydrodynamic evolution. Since multi-strange particles
are believed to kinetically decouple close to $T_c$, this suggests
that the hydrodynamic evolution in the default AZHYDRO does not
generate enough flow in the QGP. To investigate the effect of a
larger flow on HQ spectra we will also conduct Langevin simulations
with a schematic fireball whose final-state flow is given by the
empirically extracted quark spectra. The pertinent elliptic flow
exhibits slightly flatter $p_t$ dependence than in the hydrodynamic
simulation, cf.~lower panel of Fig.~\ref{fig_AZHYDRO}.  However, the
integrated quark elliptic flow of $\langle v_2 \rangle =4.99\%$ is
very close to the hydro result of $\langle v_2 \rangle =5.03\%$,
representing the benchmarks from which the heavy quarks acquire
$v_2$ through heavy-light parton interactions. However, another
20-30\% is typically built up in the hadronic evolution below $T_c$
(recall the upper panel of Fig.~\ref{fig_AZHYDRO}) which is
neglected in the present study.

\subsection{Initial Distributions of Heavy Quarks}
\label{ssec_2.4}
\begin{figure}[!t]
\hspace{4mm}
\includegraphics[width=\columnwidth
]{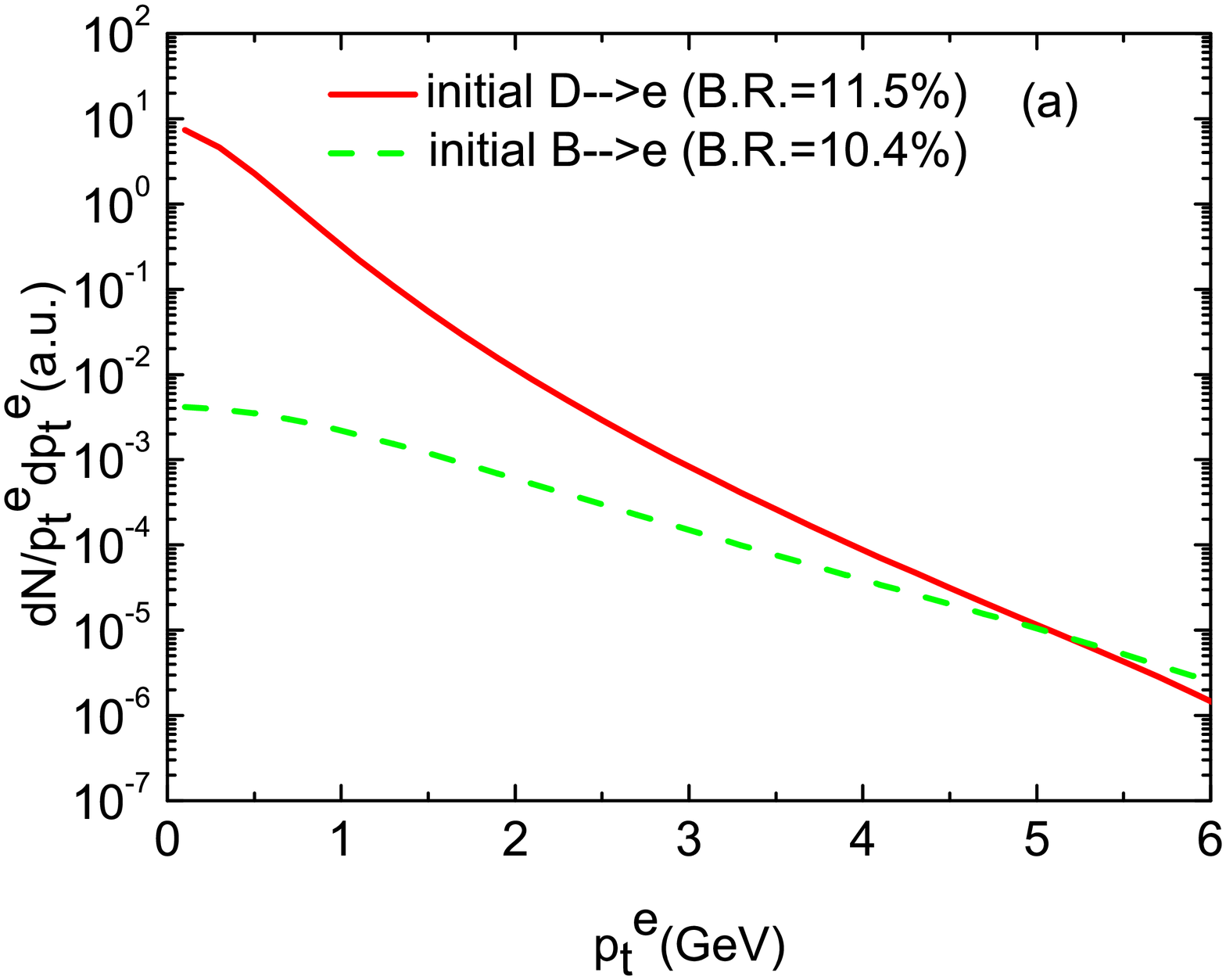}
\includegraphics[width=\columnwidth
]{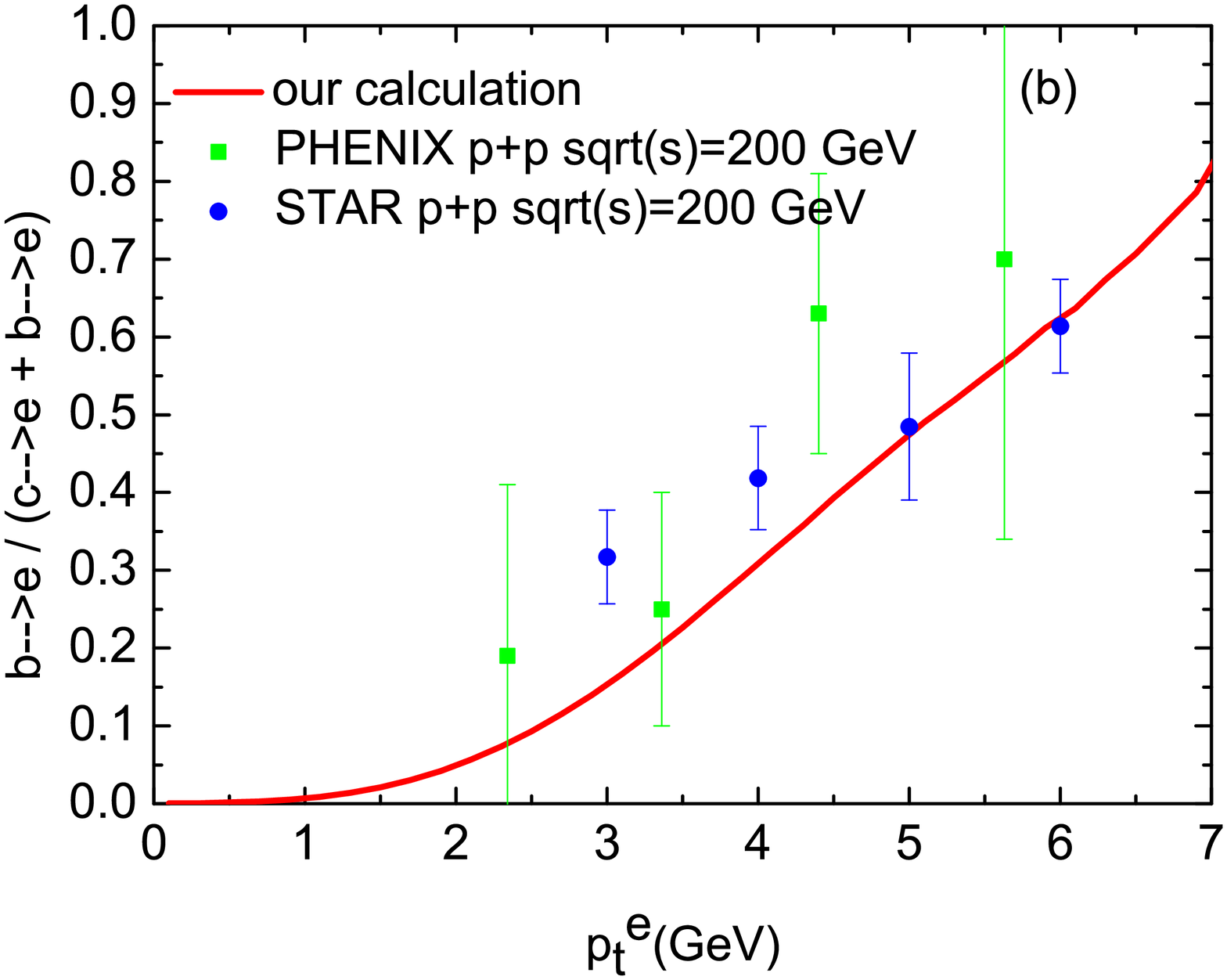} \caption{(Color online) (a) Electron
spectra from semileptonic decays of $D$- and $B$-mesons
(obtained from initial $c$- and $b$-quark spectra with
$\delta$-function fragmentation) in $p+p$ collisions at RHIC energies.
(b) Transverse-momentum dependence
of the relative contribution of electrons from $B$-mesons
to electrons from $D$+$B$ decays. The solid curve results from the
spectra in the upper panel which we adopt in our calculations; the
data are from PHENIX~\cite{Adare:2009ic} (filled squares) and from
STAR~\cite{Aggarwal:2010xp} (filled circles) for $p+p$ collisions at
$\sqrt{s}=200$\,GeV.}
\label{fig_initial-e}
\end{figure}

The number of heavy quarks produced in heavy-ion collisions is
consistent with binary nucleon-nucleon collision
scaling~\cite{Adler:2004ta}. Thus their initial spatial distribution is
expected to follow the binary collision density, $n_{\rm BC}(x,y;b)$
which we adopt in our simulations within the transverse area where the
energy density, $e(\tau_0,x,y)$, is larger than the decoupling value
$e_{\rm dec}=0.445~{\rm GeV/fm^3}$. For the initial HQ momentum
distribution, we use the same spectrum as in
Refs.~\cite{vanHees:2004gq,vanHees:2005wb}, where PYTHIA results for
charm- and bottom-quark spectra, converted into $D$ and $B$ mesons via
$\delta$-function fragmentation, were tuned to semi-leptonic
electron-decay spectra as measured in $p+p$ and $d+$Au collisions at
RHIC. This procedure leads to a bottom-to-charm cross section ratio
of $\sigma_{b\bar b}/\sigma_{c\bar c}=4.9\times10^{-3}$, and a crossing
of the electron spectra from $D$- and $B$-meson decays at
$p_t^e \approx 5$ GeV, see Fig.~\ref{fig_initial-e}. The $b/c$
cross-section ratio is within the range of pQCD
predictions~\cite{Cacciari:2005rk} and turns out to reproduce fairly
well experimental data~\cite{Adare:2009ic,Aggarwal:2010xp}
for the $p_t$-dependence of the ratio of electrons from $B$-mesons
to the sum from $D$+$B$, cf.~lower panel of Fig.~\ref{fig_initial-e}.
The $B$-meson contribution becomes sizable for $p_t^e\gsim3$\,GeV.

\subsection{Heavy-Quark Spectra and Elliptic Flow}
\label{ssec_2.5}
We now combine the ingredients as specified in the previous sections
to perform the hydro+Langevin simulation of HQ diffusion in the
QGP using the test-particle method. A vector ($\vecx_0,
\vecp_0$) in transverse phase space, representing a heavy quark, is
generated by Monte Carlo methods following the initial distributions
discussed in Sec.~\ref{ssec_2.4}. Then we follow the
trajectory of the heavy quark in phase space in equal time steps in
the lab frame. At each time step, we read off the temperature, energy
density and velocity of the fluid cell at the current HQ position,
($\tau,x,y,\eta=0$). The drag coefficient is determined by the HQ
momentum in the fluid rest frame and the temperature of the fluid
cell. The momentum of the heavy quark is updated stochastically in
the fluid rest frame according to the Langevin
rule in Eq.~(\ref{Langevinrule2}) and boosted back to the lab
frame using the fluid velocity. The HQ position is updated in the
lab frame, which can be shown to be equivalent to an
update in the fluid rest frame. Test particles that have diffused
away from $\eta=0$ to rapidity $y$ and space-time rapidity $\eta$
are redefined from the longitudinal phase-space coordinate $(\eta;y)$
to $(0;y-\eta)$ to enforce boost-invariance.

The heavy quark continues to diffuse in the QGP until the local
energy density of the fluid drops below the decoupling value, $e_{\rm
dec}=0.445~{\rm GeV/fm^3}$, corresponding to the end of the mixed
phase of the cell. At that point, we assume the heavy quark to
decouple from the fireball and mark it for hadronization.
We do not take into account a possible
local reheating if the expanding QGP phase ``swallows" again an already
hadronized heavy quark due to the increasing matter flow.
Our criterion for the decoupling of heavy quarks automatically yields
their flux across the hadronization hypersurface as
\begin{equation}
  f_Q(\tau,x,y;\vecp) p_\mu d\sigma^\mu(\tau,x,y)/E(\vecp)
  \label{eq:flux}
\end{equation}
for any area element $d\sigma^\mu(\tau,x,y)$ on that surface, in
accordance with the Cooper-Frye formalism for the hydrodynamic
freeze-out.

\begin{figure}[!t]
\hspace{4mm}
\includegraphics[width=\columnwidth
]{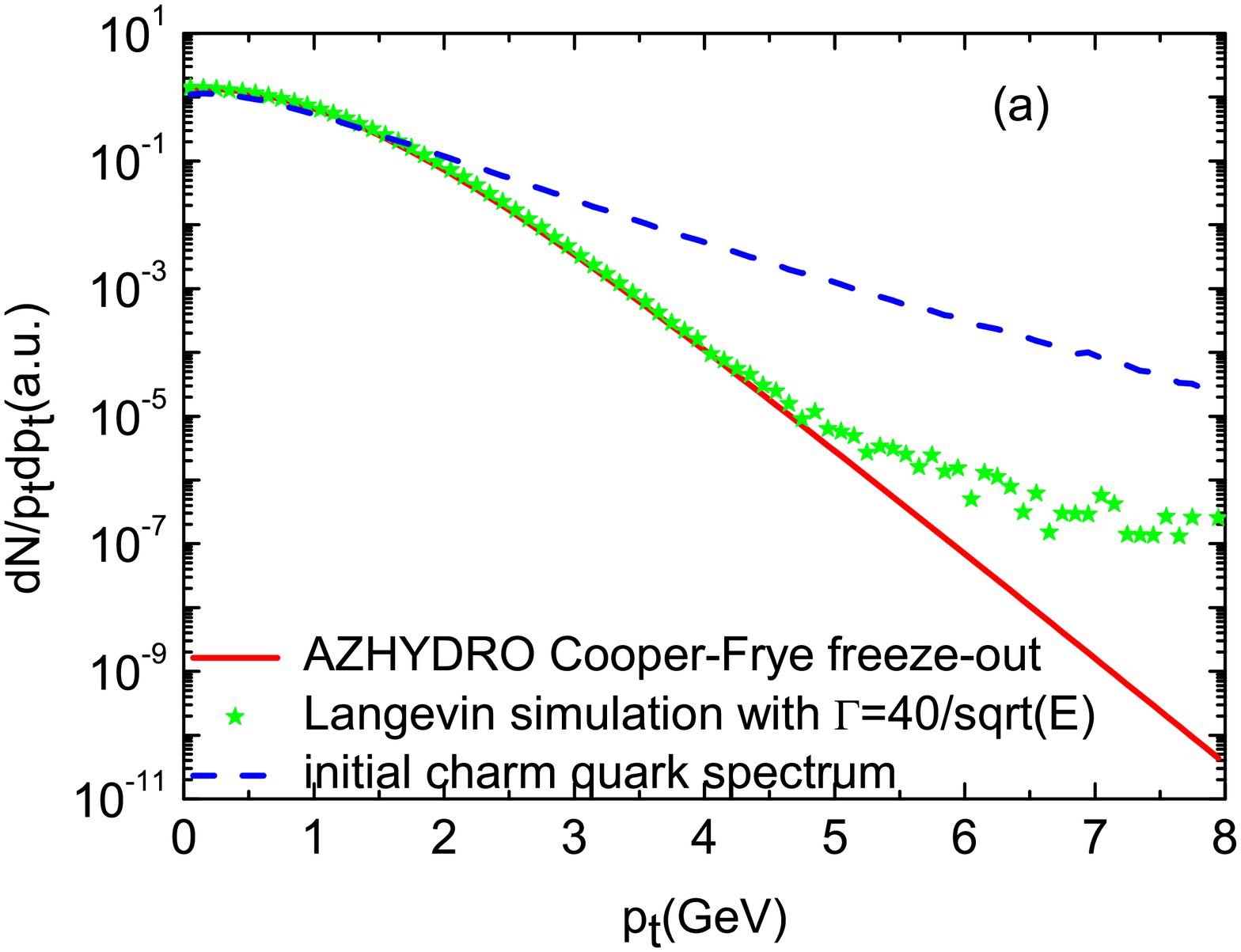}
\includegraphics[width=\columnwidth
]{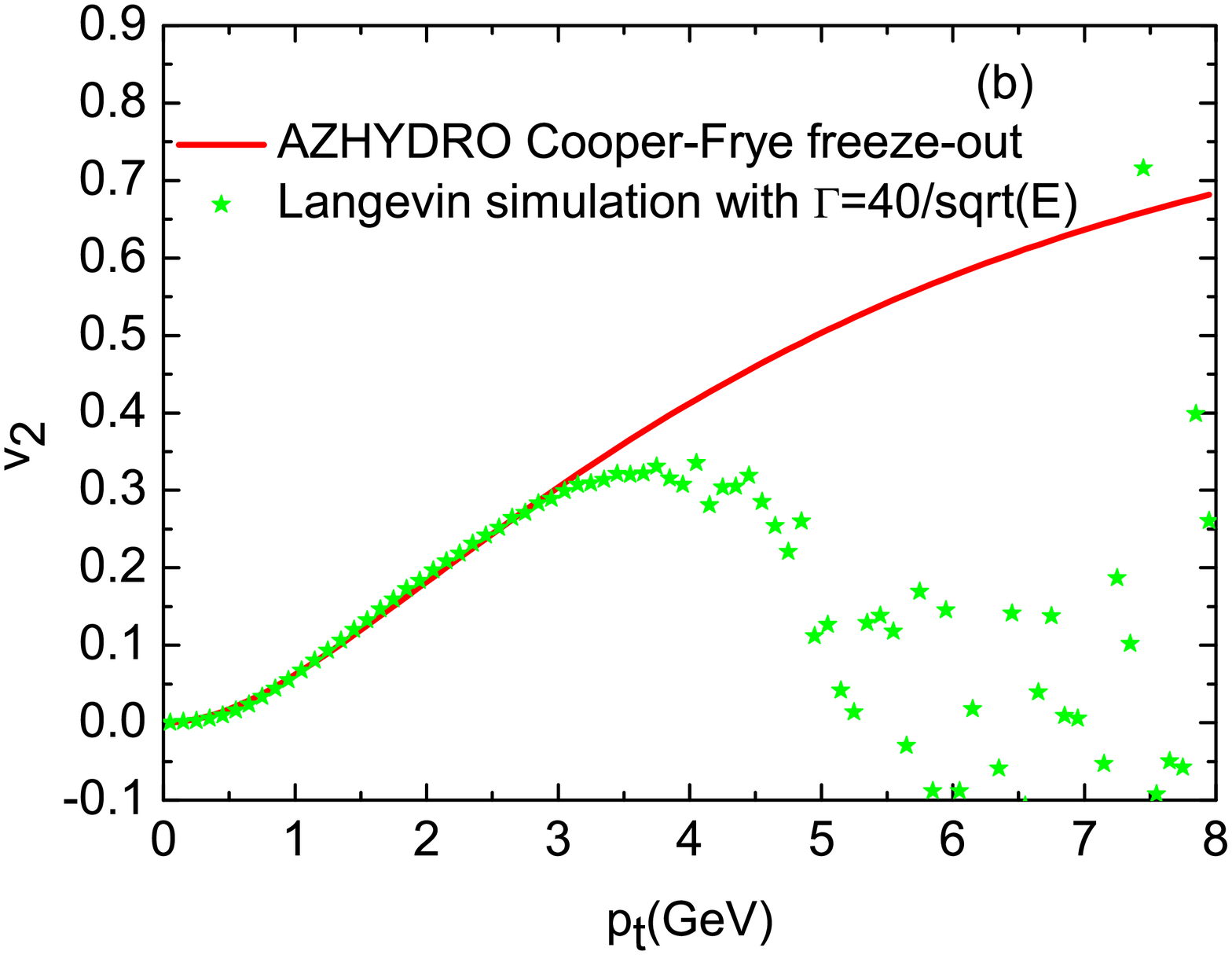} \caption{(Color online) (a) The charm
quark $p_t$-spectrum obtained from hydro+Langevin simulations with a
large drag coefficient, $\Gamma=40.0/\sqrt{E}/{\rm fm}$ (dots),
compared to the equilibrated charm-quark spectrum calculated from
the $e_{\rm dec}=0.445~{\rm GeV/fm^3}$ freeze-out hypersurface in
AZHYDRO (red solid line). The blue dashed line is the initial
charm-quark spectrum with the same total yield. (b) The same
comparison as in (a) but for the elliptic flow.}
\label{fig_c-equil}
\end{figure}
It is critical to verify that heavy quarks can reach local
equilibrium as the stationary solution~\cite{Walton:1999dy}. We have
checked the equilibrium limit with an artificially increased drag
coefficient, $\Gamma=40/\sqrt{E\,{\rm[GeV]}}/{\rm fm}$, and a
homogeneous initial spatial distribution for test particles in the
transverse plane. This specific choice for the energy dependence of
$\Gamma$ resembles the momentum dependence of the $T$-matrix based
coefficients.\footnote{We have verified that the higher-order terms
in Eq.~(\ref{Gamma-A}), which are dropped in our Langevin
simulations below, are negligible for the much smaller ``realistic"
coefficients.} The size of the numerical coefficient ($\sim$40) in
the large-$\Gamma$ case is limited by the requirement that the
numerical time-step in the Langevin process be smaller than the
inverse relaxation rate. In the upper panel of
Fig.~\ref{fig_c-equil} the Langevin charm-quark spectrum with large
coefficients is compared to the distribution from Cooper-Frye
freeze-out on the $e_{\rm dec}=0.445~{\rm GeV/fm^3}$ hypersurface in
AZHYDRO, i.e., charm quarks in complete local thermal equilibrium.
We have adopted a charm-quark mass of $m_c=1.8$\,GeV, corresponding
to the in-medium mass at $T_c=165$\,MeV in our
simulations~\cite{Riek:2010fk}. The spectra agree well up to
$p_t\simeq 4.0-4.5~{\rm GeV}$. The deviation at higher $p_t$ is due
to surface emission of charm quarks with large velocities which
escape the active (i.e.\ $e(\tau,x,y)\geq e_{\rm dec}$) part of the
fireball at the earliest times; roughly 1\% of heavy quarks at a
given high $p_t$ do not suffer collisions, corresponding to the
factor $\sim$100 suppression of spectra from the Langevin simulation
relative to the initial distribution at large $p_t$. A matching
picture is observed for the elliptic flow (lower panel in
Fig.~\ref{fig_c-equil}): at low $p_t$ the $v_2$ of the
hydro+large-$\Gamma$-Langevin simulation follows the $v_2$ of
equilibrated charm quarks, while it breaks away and oscillates
around zero for large $p_t$ (deviations set in slightly earlier than
for the inclusive $p_t$ spectra, presumably since $v_2$ is a more
differential and thus more ``fragile" quantity).

\begin{figure}[!t]
\hspace{4mm}
\includegraphics[width=\columnwidth
]{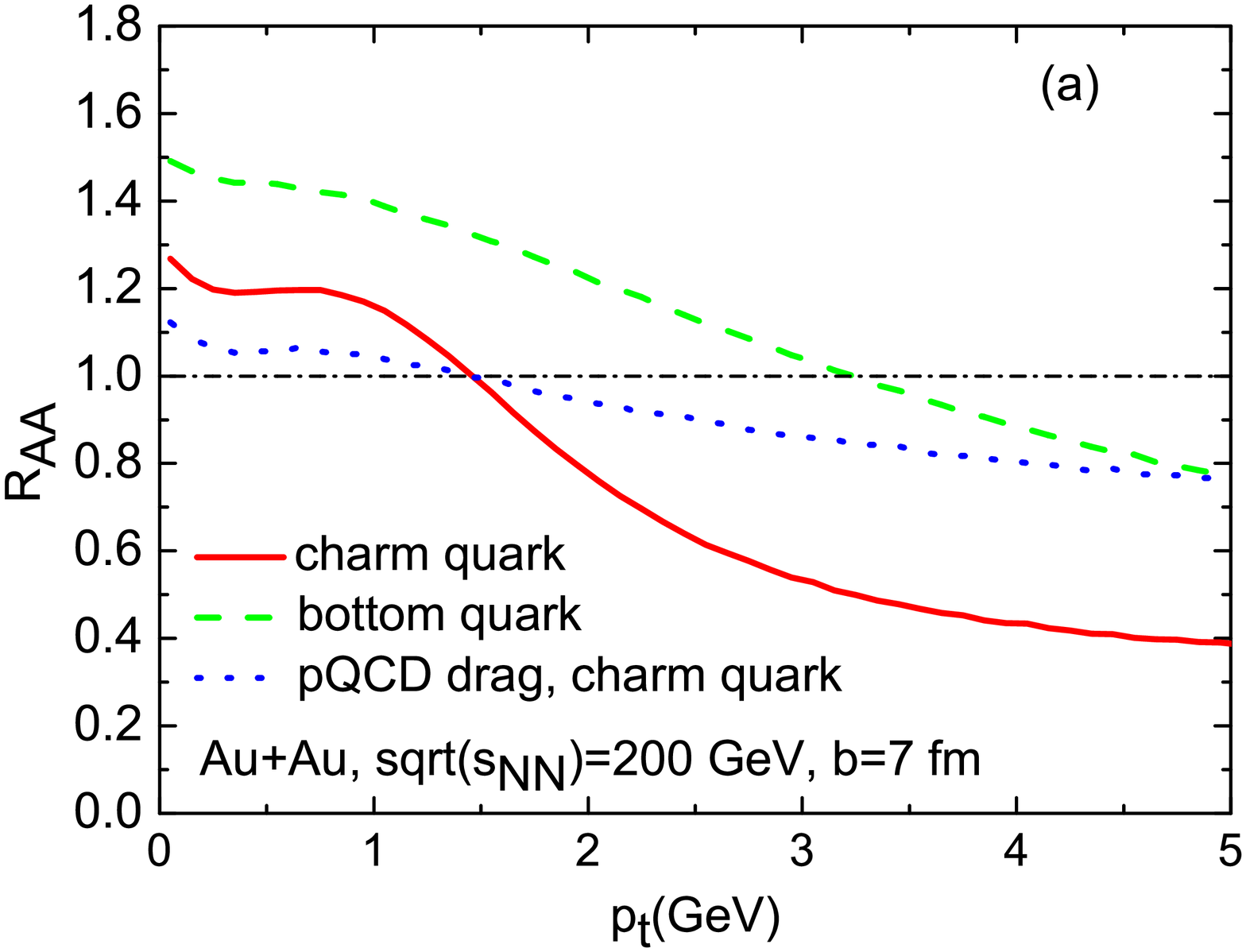}
\includegraphics[width=\columnwidth
]{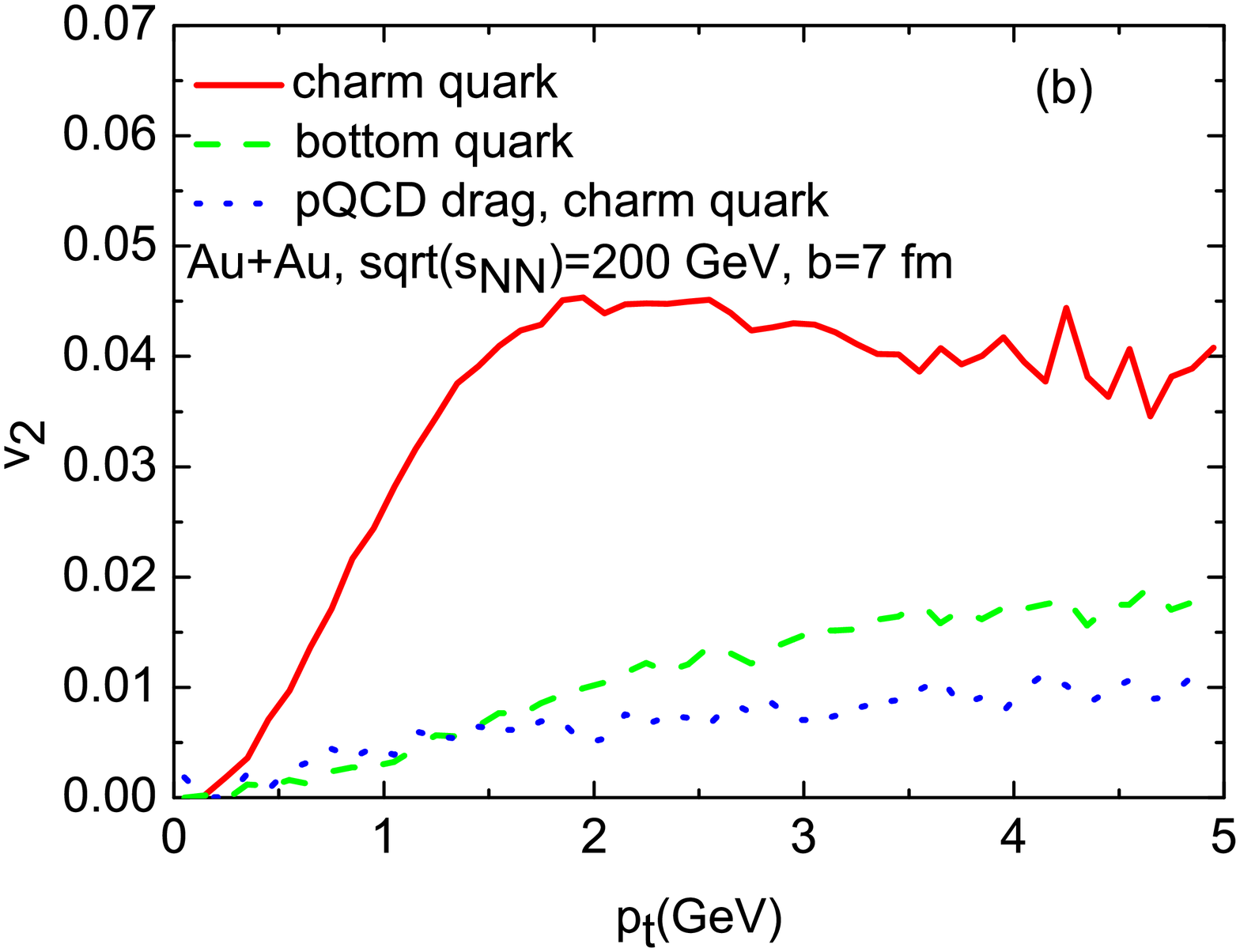}
\caption{(Color online) Nuclear modification factor (upper panel) and
elliptic flow (lower panel) of charm (red solid line) and bottom quarks
(green dashed line) at hadronization obtained from hydro+Langevin
simulations for $b$=7\,fm Au+Au collisions at RHIC energy, using transport
coefficients from the heavy-light quark $T$-matrix plus a pQCD HQ-gluon
contribution. For comparison charm-quark results are shown with
coefficients using only LO pQCD scattering off gluons light quarks (blue
dotted line).}
\label{fig_hq-raa-v2}
\end{figure}
Next we turn to the results of our simulations under ``realistic"
conditions, using the transport coefficients and initial distributions
outlined above, together with temperature-dependent in-medium HQ
masses~\cite{Riek:2010fk}. As usual, the modifications of the HQ spectra
in the medium are quantified by the nuclear modification factor and
elliptic flow,
\begin{align}
\label{RAAv2}
  R_{AA}(p_t,y) &= \frac{\frac{dN_{\rm AA}}{dp_tdy}}{N_{\rm coll} \frac{dN_{\rm
  pp}}{dp_tdy}} \ ,\\
  v_2(p_t,y) &= \frac{\int d\phi\frac{dN_{\rm AA}}{dp_td\phi
  dy}\cos(2\phi)}{\int d\phi\frac{dN_{\rm AA}}{dp_td\phi dy}}  \ ,
\end{align}
respectively, where $N_{\rm coll}$ is the estimated number of binary
nucleon-nucleon collisions for the centrality bin under consideration.
In Fig.~\ref{fig_hq-raa-v2} we display the charm- and bottom-quark
$R_{\rm AA}$ and $v_2$ at the end of the mixed phase as obtained
from hydro+Langevin simulations in semi-central Au+Au collisions
($b$=7\,fm). The approach toward thermalization induces a depletion
of heavy quarks at large $p_t$ (quenching) and an enhancement at low
$p_t$ enforced by HQ number conservation. At $p_t\simeq5$\,GeV, the
charm-quark quenching reaches down to $\sim$0.4 while bottom quarks
are much less affected, with $R_{\rm AA}$($p_t$=5\,GeV)$\simeq0.8$.
Note that at the same $p_t$ the Lorentz-$\gamma$ of bottom quarks is
significantly smaller than for charm. Radiative contributions to HQ
transport are estimated to become competitive with elastic
scattering once the non-perturbative effects are suppressed, i.e.,
above $p_t\simeq$~4-5\,GeV for charm quarks~\cite{Rapp:2009my}
(recall Fig.~\ref{fig_Ap}). When using a drag coefficient from pQCD
elastic scattering only (including both quarks and gluons with
$\alpha_s=0.4$), the quenching is weaker by about a factor of
$\sim$3~\cite{vanHees:2005wb}. Similar features are found in the
elliptic flow coefficient, which for $c$-quarks first increases
approximately linearly before leveling off at about 4.5\%,
characterizing a transition from a quasi-thermal to a kinetic
regime.

Previous calculations employing a thermal-fireball model for the
medium evolution, using drag coefficients for non-perturbative
elastic scattering~\cite{vanHees:2004gq} of comparable magnitude as
in our calculation, have found significantly larger values for the
maximal charm-quark $v_2$ of around 7.5\%~\cite{vanHees:2005wb}.
Part of this difference originates from the larger ``intrinsic"
$v_2$ in the fireball medium which has been adjusted to the
empirically observed hadron-$v_2$ of 5.5-6\%.  Since the diffusion
coefficient of charm ($D$-mesons) in the hadronic phase is not
negligible~\cite{He:2011yi}, the HF $v_2$ in the present study
should be considered as a lower bound. Another source of uncertainty
derives from the freeze-out prescription and the associated
realization of the HQ Langevin process (Cooper-Frye in the hydro
evolution vs. Milekhin-like in some fireball
calculations)~\cite{Gossiaux:2011ea}.

\section{Heavy-Quark Hadronization}
\label{sec_hadronization}
The bulk matter in a hydrodynamic simulation can be evolved through
a phase transition (here QGP to hadronic matter) solely by
specifying the equation of state of the medium. However, the HQ
spectra resulting from the Langevin simulations through the QGP are,
in general, not in full equilibrium with the bulk medium and thus
require a microscopic hadronization mechanism to enable the
calculation of HF observables. We will carry this out at the end of
the mixed phase, represented by the hypersurface defined by the
critical energy density of the hadronic phase in the hydrodynamic
simulation. For simplicity we focus on the formation of $D$-
and $B$-mesons neglecting HF baryons and hidden heavy flavor (both
of which have been found to give small contributions to the total HF
content of the hadronic phase~\cite{vanHees:2005wb}). Two
microscopic hadronization mechanisms have been considered in
heavy-ion physics to date: independent fragmentation of partons and
coalescence of quarks. The former is appropriate for large-momentum
partons emerging directly from initial hard processes,
with phenomenological fragmentation functions simulating vacuum gluon
radiation and color neutralization.
Coalescence, on the other hand, is believed to dominate in the
low-momentum regime where partons are abundant in phase-space in
heavy-ion~\cite{Fries:2003vb,Greco:2003mm,Fries:2008hs} and even
in elementary hadronic reactions~\cite{Rapp:2003wn,Hwa:2005}.

Several previous studies of HQ diffusion in heavy-ion collisions
have neglected coalescence
processes~\cite{Akamatsu:2008ge,Das:2009vy,Alberico:2011zy,Uphoff:2010sh},
thus limiting the applicability of HF observables to high momenta.
The formation time of heavy quarks is comparatively short, and
thereafter their virtuality is small, governed by interactions with
the medium with modest momentum transfers. Hence, fragmentation is
not effective. In the Langevin simulations of
Refs.~\cite{vanHees:2005wb,vanHees:2007me,Gossiaux:2008jv},
heavy-light quark recombination has been accounted
for~\cite{Greco:2003vf} and found to be important for increasing
{\em both} the elliptic flow and the nuclear modification factor of
the resulting $D$-meson spectra. The coalescence formalism was based
on the widely used instantaneous
approximation~\cite{Fries:2003vb,Greco:2003mm} which, however, does
not conserve energy in the $2\to1$ hadron formation process. A
related problem is the lack of a well-defined equilibrium limit for
the hadron distributions. Together, both features imply appreciable
uncertainties in calculating HF observables in the low-$p_T$ region
(albeit suppressed compared to light-quark coalescence by a mass
ratio $(m_q/m_c)$). To improve the coalescence description and
achieve consistency with local kinetic equilibrium, we here employ
the resonance recombination model (RRM) implemented on the
hydrodynamic hadronization surface.

\subsection{Resonance Recombination at the Hadronization Hypersurface}
\label{ssec_rrm}
In the RRM the hadronization of constituent quarks is treated via
resonance scattering within a Boltzmann transport
equation~\cite{Ravagli:2007xx}. For scattering rates which are large
compared to the inverse hadronization time, $\Gamma_{\rm res}\gg
1/\tau_{\rm had}$, equilibrium quark distribution functions in a
flowing medium are converted into equilibrium meson spectra with the
same flow properties, including elliptic anisotropies with
space-momentum correlations characteristic for a hydrodynamically
expanding source~\cite{Ravagli:2008rt,He:2010vw}. The RRM has been
employed previously to investigate kinetic-energy and
constituent-quark number scaling~\cite{Ravagli:2008rt}, and to
extract empirical quark distribution functions of the bulk medium at
hadronization at RHIC~\cite{He:2010vw}.

The RRM is consistent with the heavy-light Feshbach resonance
formation found in the $T$-matrix calculation of the HQ thermal
relaxation rate (see Section~\ref{ssec_2.2}). It reiterates the
important role played by resonance correlations in our work. As the
temperature drops towards $T_c$, the resonance correlations in the
heavy-light quark $T$-matrix strengthen (recall Fig.~\ref{fig_Tmat})
and thus naturally merge into heavy-light quark recombination
processes. When implementing the latter via a Breit-Wigner ansatz
one obtains the HF meson distribution from the asymptotic solution
of the Boltzmann equation as~\cite{Ravagli:2007xx}
\begin{multline}
f_M^{\mathrm{asymp}}(\vecx,\vecp)=\frac{E_M(\vecp)}{m_M\Gamma_M}
\int\frac{d^3p_1d^3p_2}{(2\pi)^6}f_Q(\vecx,\vecp_1)
\\
\times f_{\bar q}(\vecx,\vecp_2) \ \sigma(s) \
v_{\rm rel}(\vecp_1,\vecp_2) \ \delta^{(3)}(\vecp -\vecp_1 -\vecp_2) \ ,
\label{rrm}
\end{multline}
where $f_{Q,q,M}$ are equal-time phase-space distributions of heavy
quarks, light quarks, mesons, respectively, $v_{\rm rel}$ is the
relative velocity of the recombining heavy and light quarks, and
$m_M$ and $\Gamma_M$ are the mass and width of the meson resonance
~\cite{Ravagli:2007xx,Ravagli:2008rt,He:2010vw}. In the
calculations below we employ masses and widths compatible with the
$T$-matrix calculation of Ref.~\cite{Riek:2010fk}, extrapolated to
$T_c=165$~MeV with $m_c=1.8$\,GeV, $m_q=0.35$\,GeV, $m_D=2.25$ GeV
and $\Gamma_D=0.1$\,GeV.

Energy conservation and detailed balance in RRM ensure an
equilibrium mapping between the distributions of quarks and formed
mesons~\cite{Ravagli:2008rt,He:2010vw}. We verify this in the present
case of a non-trivial freeze-out hypersurface given by AZHYDRO. We use
local charm- and light-quark equilibrium phase-space distributions,
$f(p,x)=e^{-p\cdot u(x)/T}$, with fluid velocities given by AZHYDRO at
the end of the mixed phase, then apply resonance recombination,
Eq.~(\ref{rrm}), locally (for each cell) to obtain the local meson
phase-space distribution $f_M(\tau,x,y;\vecp)$. Finally we calculate the
current across the hypersurface and sum over all fluid cells on the
$e_{\rm dec}=0.445~{\rm GeV/fm^3}$ freeze-out hypersurface,
\begin{equation}
\label{CF}
  \frac{dN}{p_Tdp_Td\phi dy}= \int_{\Sigma}\frac{p_\mu
  d\sigma^\mu(\tau,x,y)}{(2\pi)^3} f_M (\tau,x,y;\vecp) \ .
\end{equation}

\begin{figure}[!t]
\hspace{4mm}
\includegraphics[width=\columnwidth
]{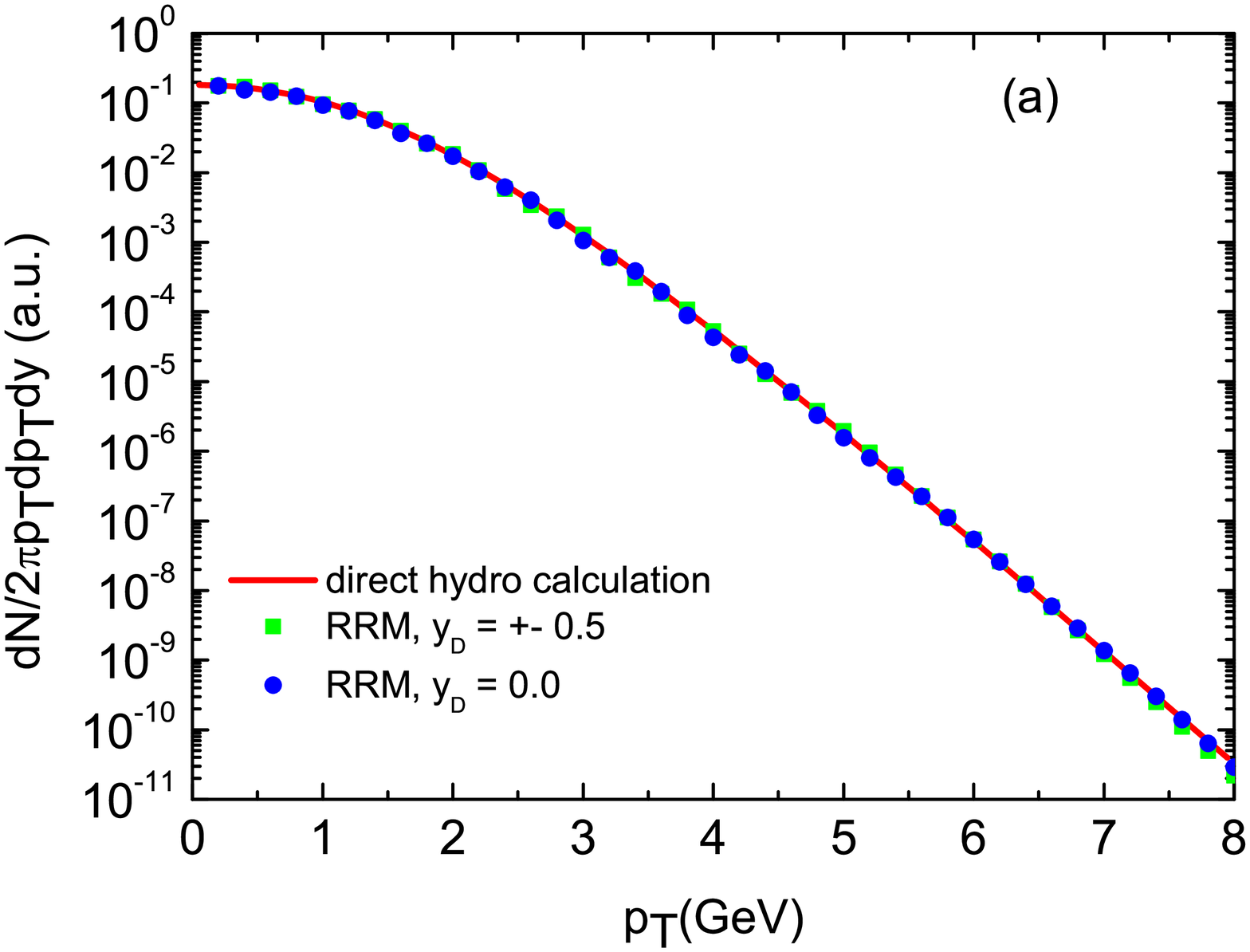}
\includegraphics[width=\columnwidth
]{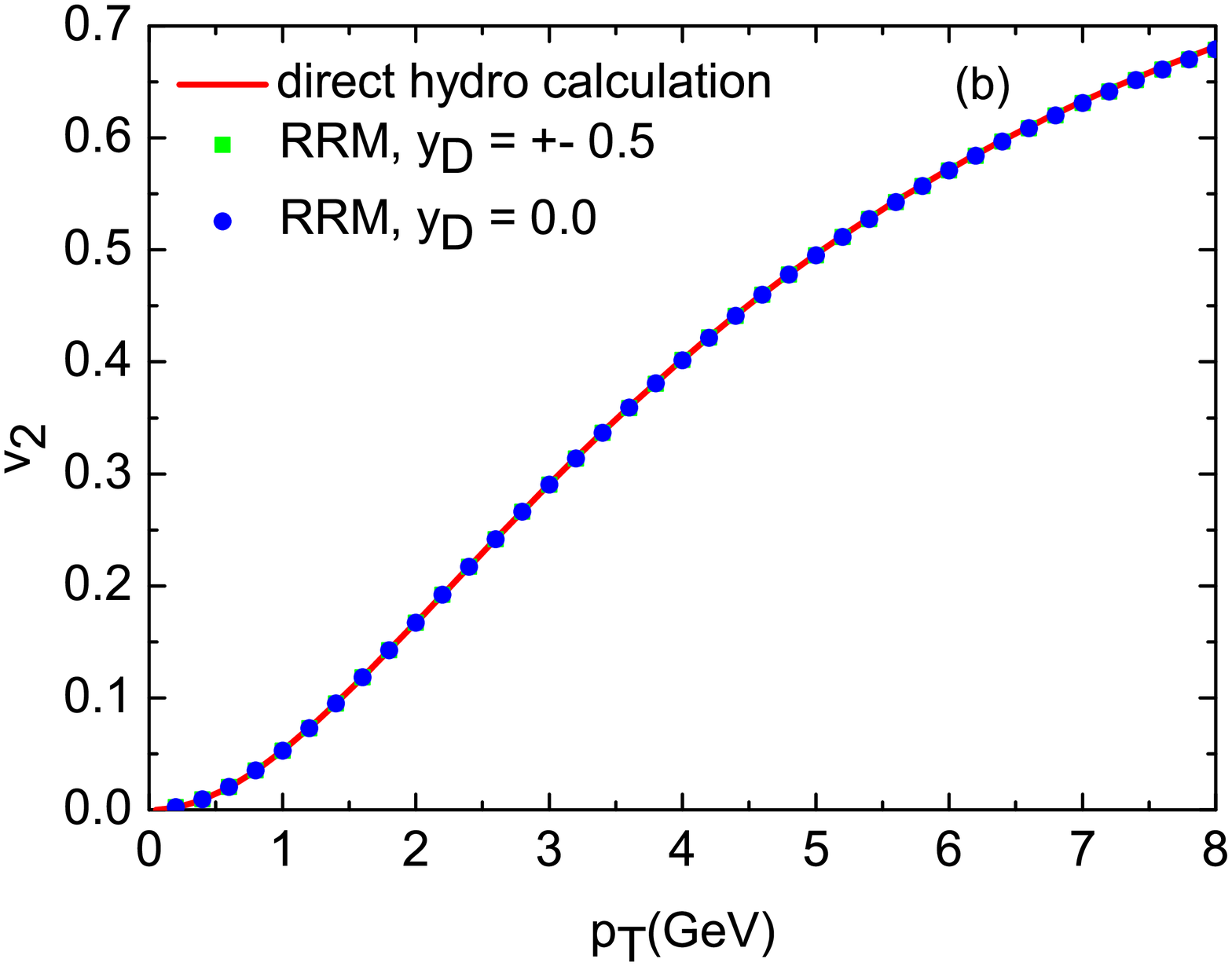}
\caption{(Color online) (a) $D$-meson $p_T$-spectrum calculated with
RRM on the AZHYDRO hadronization hypersurface (circles), compared to
a direct calculation from AZHYDRO using the Cooper-Frye formula on
the same hypersurface (solid line). The $D$-meson spectra at different
rapidities ($y_D=0.0$ and $y_D=\pm 0.5$) calculated from RRM agree with
each other. (b) The same comparison for the elliptic flow of $D$ mesons.}
\label{fig_hydroRRM}
\end{figure}

In Fig.~\ref{fig_hydroRRM} we compare the resulting $D$-meson spectrum
and $v_2$ with a calculation directly from hydro using $D$-meson Cooper-Fry
freeze-out on the same hypersurface ($e_{\rm dec}=0.445~{\rm GeV/fm^3}$).
The close agreement of the two calculations verifies the mapping
between the equilibrium quark and meson distributions in RRM,
including the full space-momentum correlations encoded in the AZHYDRO
flow field. Longitudinal boost invariance of AZHYDRO
is preserved by RRM as well, as observed from the independence of the
$D$-meson spectra on rapidity within our accuracy.

\begin{figure}[!t]
\hspace{4mm}
\includegraphics[width=\columnwidth
]{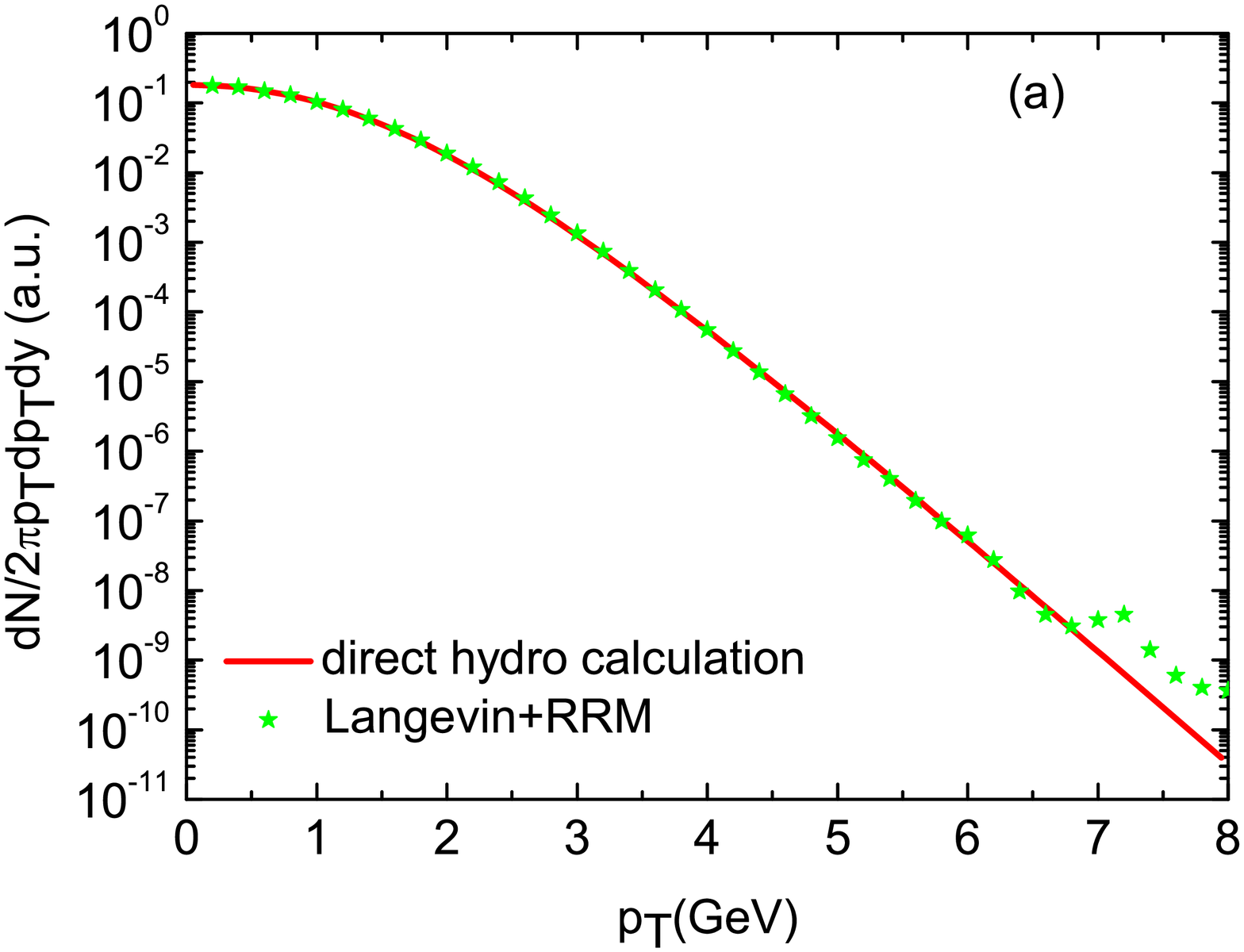}
\includegraphics[width=\columnwidth
]{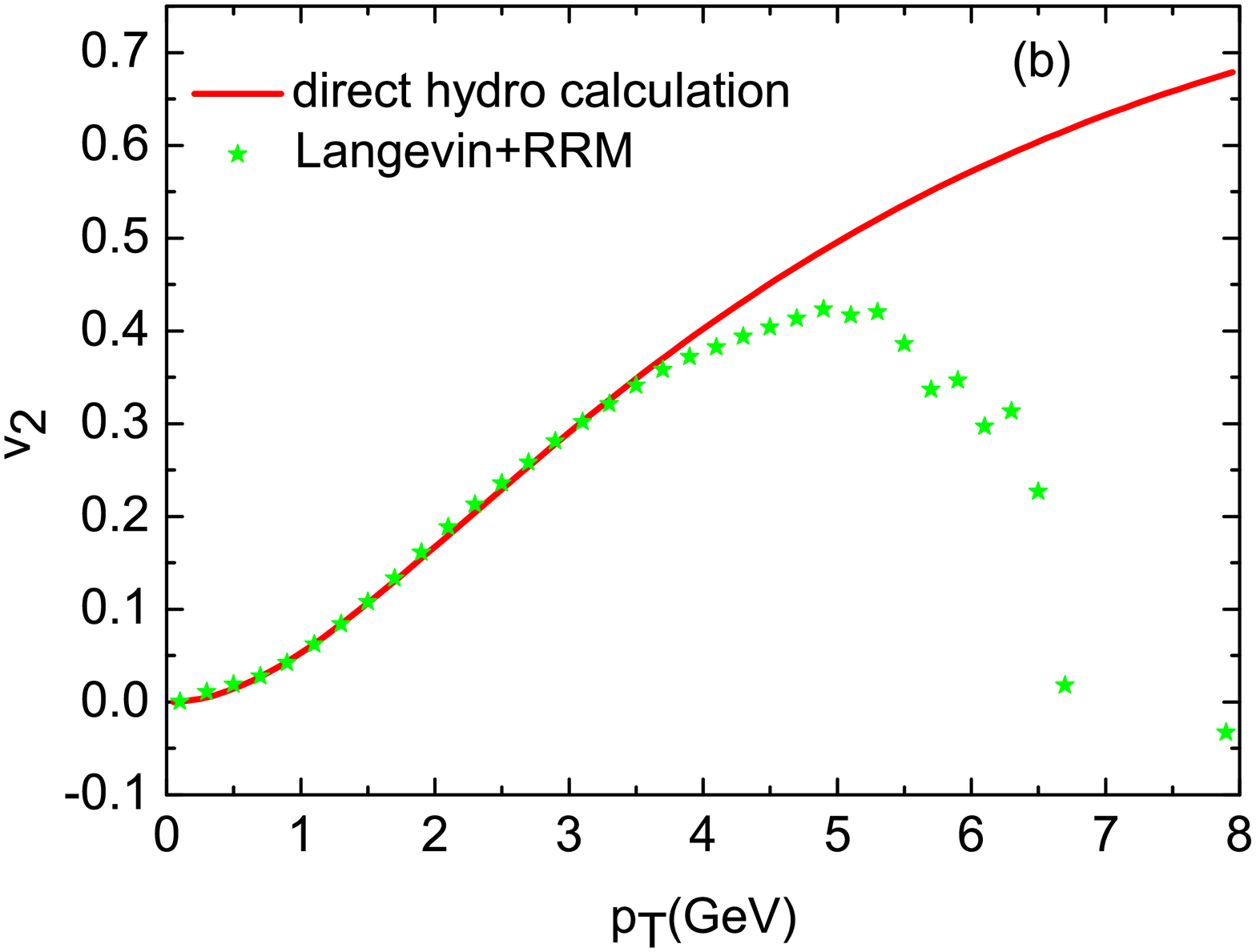}
\caption{(Color online) (a) $D$-meson $p_T$-spectrum (stars)
calculated from RRM on the hadronization hypersurface applied to
charm-quark spectra from hydro+Langevin simulations in the large-drag
coefficient limit (corresponding to Fig.~\ref{fig_c-equil}). It is compared to
the $D$-meson spectrum directly calculated from AZHYDRO on the same hypersurface.
(b) Same as in panel (a) but for $D$-meson elliptic flow.}
\label{LangeviRRMequilibrium}
\end{figure}

The next step is to extend our approach to hadronize off-equilibrium
quark distribution functions emerging from our HQ Langevin simulations.
In order to couple the RRM to a HQ test particle freezing out from the
hydro-Langevin simulation with momentum $\vecp_{\rm dec}$ and coordinate
$\vecx_{\rm dec}$, we represent the corresponding local equal-time HQ
phase-space distribution on the hadronization hypersurface, $f_Q$ from
Eq.~(\ref{eq:flux}), by a $\delta$-function,
$\delta^3(\vecx -\vecx_{\rm dec})\delta^3(\vecp-\vecp_{\rm dec})$ at the
hadronization time $\tau(x_{\rm dec},y_{\rm dec})$. As before, the
light-quark phase space density $f_q$ at this point is taken to be the
equilibrium distribution $e^{-p\cdot u/T}$ at the local temperature and
flow, 
and we can apply Eq.~(\ref{rrm}) to obtain the phase space density
$f_M$ of heavy mesons test particle by test particle. Finally, the
spectrum of heavy mesons follows from Eq.~(\ref{CF}) as a sum over
test particles. To check our procedure we first apply it to the
Langevin output in the large-coefficient limit for charm quarks
which -- as discussed in Sec.~\ref{ssec_2.5} -- follows the
equilibrium distribution up to $p_t\simeq$~4\,GeV.
Figure~\ref{LangeviRRMequilibrium} shows the comparison between the
$D$-meson spectrum and $v_2$ calculated in this way and the
$D$-mesons from a direct AZHYDRO calculation. Compared to
Fig.~\ref{fig_c-equil}, the agreement of the hydro+Langevin+RRM
calculation with the direct hydro calculation extends to slightly
larger $p_T$ since the recombination essentially acts as an
additional heavy-light interaction when forming $D$-mesons. This
connection is particularly transparent if the interaction used in
the diffusion process is the same as in the recombination process,
i.e., a resonance interaction in our approach. We conclude that the
equilibrium limit is well under control in this framework.

\subsection{Hadronization: Coalescence vs Fragmentation}
\label{ssec_coal-frag}
After establishing a coalescence formalism for hadronization of
off-equilibrium HQ distributions on an arbitrary hypersurface it
remains to couple this contribution with the standard fragmentation
mechanism representing the large-$p_t$ limit where the phase-space
density of light partons vanishes (vacuum limit). In most previous
works the coalescence probability has been evaluated in an
instantaneous approximation which did not allow for a full control
over its absolute magnitude. Here, instead, we use a dynamic
criterion which directly follows from the RRM formalism; it is based
on the HQ scattering rate which is derived from the same
interactions as used in the diffusion calculations\footnote{See
Refs.~\cite{vanHees:2007me,He:2011yi} for a recent discussion of the
relation between the two in the HF context}. We thus make the
following ansatz for the HQ coalescence probability in the fluid
rest frame:
\begin{equation}
  P_{\rm coal}(p) = \Delta\tau_{\rm res} \ \Gamma_Q^{\rm res}(p) \ ,
\label{Pcoal}
\end{equation}
which is Lorentz invariant. In Eq.~(\ref{Pcoal}), the scattering
rate, $\Gamma_Q^{\rm res}= n_{q} \langle \sigma_{qQ}^{\rm res} \
v_{\rm{rel}}\rangle$, refers to the resonant part of the $Qq$ cross
section, $\sigma_{qQ}^{\rm res}$ (or $T$-matrix), and thus
represents the rate for hadron formation ($n_{q}$: light-quark
density, $v_{\rm rel}$: relative velocity). The time interval
$\Delta\tau_{\rm res}$ characterizes the window in the dynamic
medium evolution during which resonance states exist; typically,
this corresponds to the duration of the hadronization transition,
i.e., the ``mixed phase", or even longer depending on whether
``pre-resonance" states can be formed above $T_c$. Of course, if the
product $\Delta\tau_{\rm res}~\Gamma_Q^{\rm res}$ exceeds one,
$P_{\rm coal}$ should be put to one, corresponding to the
equilibrium limit (more accurately, one could apply an exponential
relaxation, but in view of the practical uncertainties in the values
for $\Delta\tau_{\rm res}$ and $\Gamma_Q^{\rm res}(p)$ this is
currently not warranted). We emphasize that this procedure provides
an absolute normalization of the coalescence contribution,
consistent with the (unique) equilibrium limit. Since the resonance
formation rate naturally diminishes with increasing HQ momentum (the
phase-space density of quarks from the thermal bath to match the
resonance mass decreases), one obtains an increasing fraction,
$P_{\rm frag}(p) = 1- P_{\rm coal}(p)$, of heavy quarks undergoing
independent fragmentation, recovering the vacuum limit.

In practice, for our calculations reported below, we evaluate
Eq.~(\ref{Pcoal}) as follows. For the HQ scattering rate we employ a
Breit-Wigner cross section which is consistent with our heavy-light
$T$-matrix (cf.~Sec.~\ref{ssec_2.2}) and approximately reproduces
the color-singlet contribution to the HQ thermal relaxation rate at
$T_c$; the pertinent meson width of $\Gamma_M$=0.4\,GeV is larger
than the one used in the RRM expression, Eq.~(\ref{rrm}), but the
resulting meson spectra and elliptic flow are rather insensitive to
this quantity~\cite{Ravagli:2007xx,Ravagli:2008rt} (we neglect
resonant diquark contributions since we only consider color-singlet
scattering relevant for $D$- and $B$-meson formation). With this
cross section we calculate the HQ scattering rate in the fluid rest
frame. We typically find $\Gamma_c^{\rm res}\approx$~0.1\,GeV for
charm quarks at vanishing momentum (similar for bottom quarks),
consistent with Refs.~\cite{Riek:2010fk,Riek:2010py} (about half of
the total HQ width of $\sim$0.2\,GeV calculated in these works is
due to the color-singlet part).

The HQ scattering rate is then boosted to the lab frame at the end
of Langevin simulation (mixed phase) and expressed as a function of
HQ transverse momentum. For simplicity, we have chosen to apply
$P_{\rm coal}$ not test-particle-by-test-particle but averaged over
the spatial dependence in the fireball, i.e., as a function of $p_t$
only. We have checked that the explicit inclusion of space-momentum
correlations leads to very similar results for heavy-meson spectra
and $v_2$. For the macroscopic time duration of resonance formation
in the medium evolution, we adopt a conservative estimate of
$\Delta\tau_{\rm res}\simeq$~2\,fm/$c$ (in the lab frame), amounting
to $P_{\rm coal}(p_t\to0)\to1$; this is shorter than the duration of
the mixed phase in the hydro evolution, $\Delta\tau_{\rm
mix}\simeq$~3-4\,fm/$c$ (roughly corresponding to the dip structure
in the upper panel of Fig.~\ref{fig_AZHYDRO}), and thus constitutes
a lower limit of the coalescence contribution.

\begin{figure}[!t]
\hspace{4mm}
\includegraphics[width=\columnwidth
]{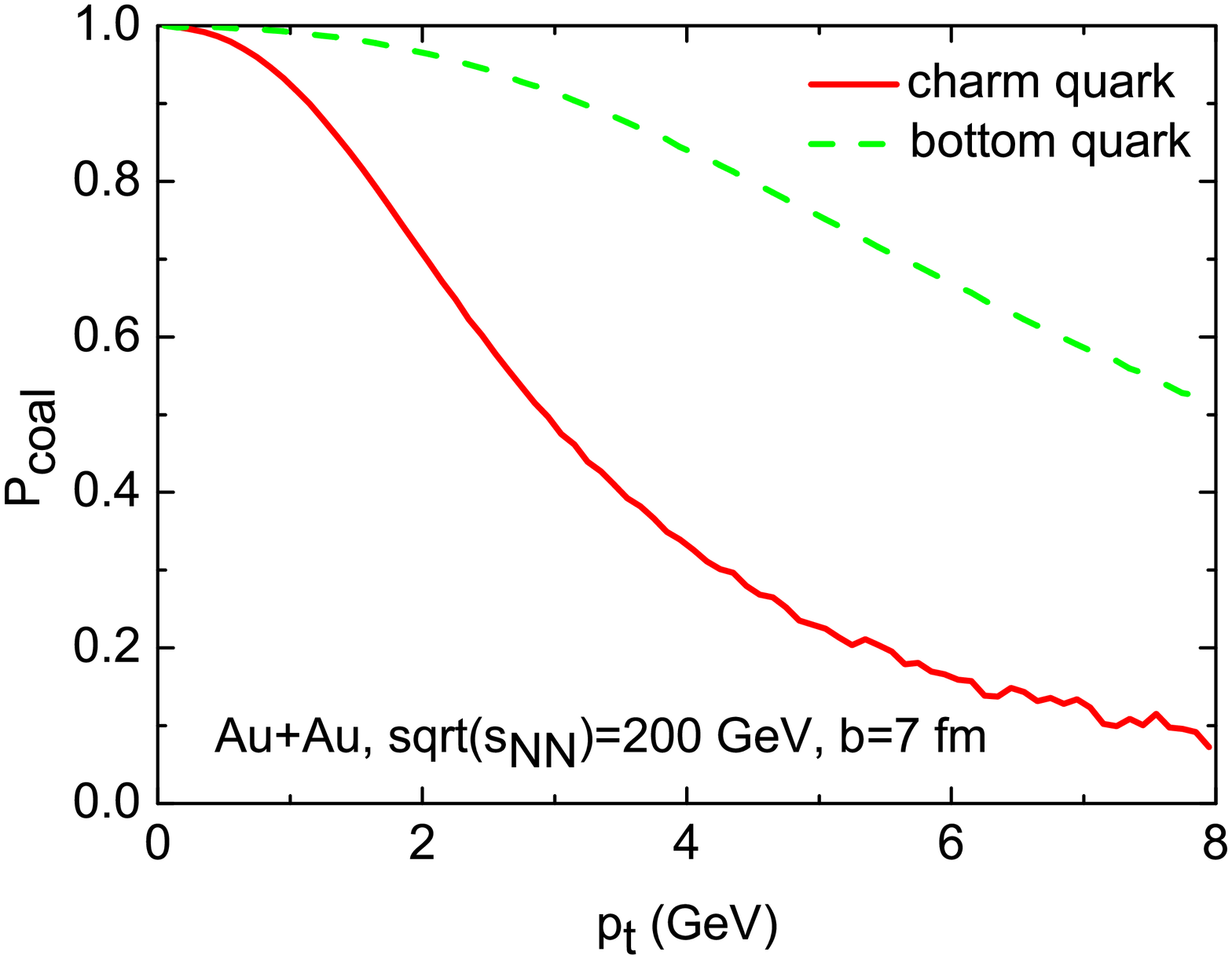}\caption{(Color online) Charm- and bottom-quark
coalescence probability as a function of lab-frame $p_t$ for
semi-central Au+Au collisions at RHIC. }
\label{fig_pcoal}
\end{figure}

In Fig.~\ref{fig_pcoal} we show the coalescence probability,
$P_{\rm coal}(p_t)$, for charm and bottom quarks in the lab frame,
averaged over all test particles with a given $p_t$ in semi-central
($b$=7\,fm) Au+Au collisions at RHIC.
At a given $p_t$, bottom quarks have smaller velocities than
charm quarks and thus more easily find a
light-quark partner to recombine with. Consequently, the $b$-quark
coalescence probability drops slower compared to $c$ quarks.


Let us summarize our procedure for HQ hadronization: for each HQ test particle
we determine a heavy-meson spectrum for both recombination and fragmentation
components, by applying either RRM or a fragmentation function
$D_Q(z) = \delta(z-1)$. We then sum over test particles using respective
weights $P_{\rm coal} (p_t)$ and $1-P_{\rm coal}(p_t)$,
yielding the full coalescence+fragmentation meson spectrum. Thus, the
total HF meson spectrum can then be written as
\begin{equation}
  \frac{dN_M^{\rm total}}{p_Tdp_Td\phi dy} = \frac{dN_M^{\rm
  coal}}{p_Tdp_Td\phi dy} + \frac{dN_M^{\rm frag}}{p_Tdp_Td\phi dy} \ .
  \label{eq:hfsum}
  \end{equation}
Note that we determine the absolute normalization of the spectrum
by requiring the conservation of HF quantum numbers. Since our
fragmentation function automatically conserves HF number we only need
to impose the normalization on the recombination contribution to the
spectrum. We recall that the
formation of heavy quarkonia and heavy baryons has been neglected.

\subsection{Numerical results for $D$ and $B$ mesons}
\label{ssec_DB}
\begin{figure}[!t]
\hspace{4mm}
\includegraphics[width=\columnwidth
]{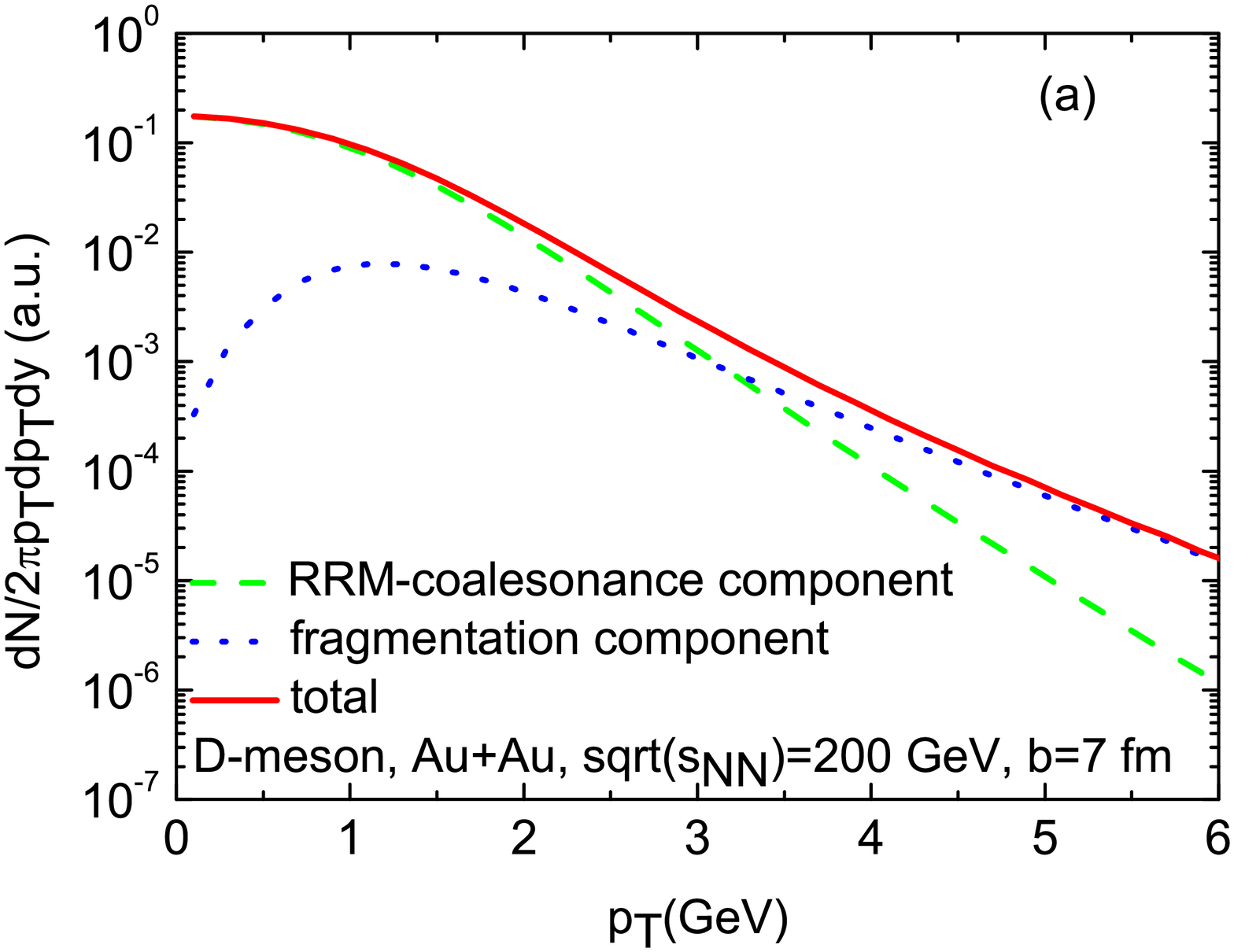}
\includegraphics[width=\columnwidth
]{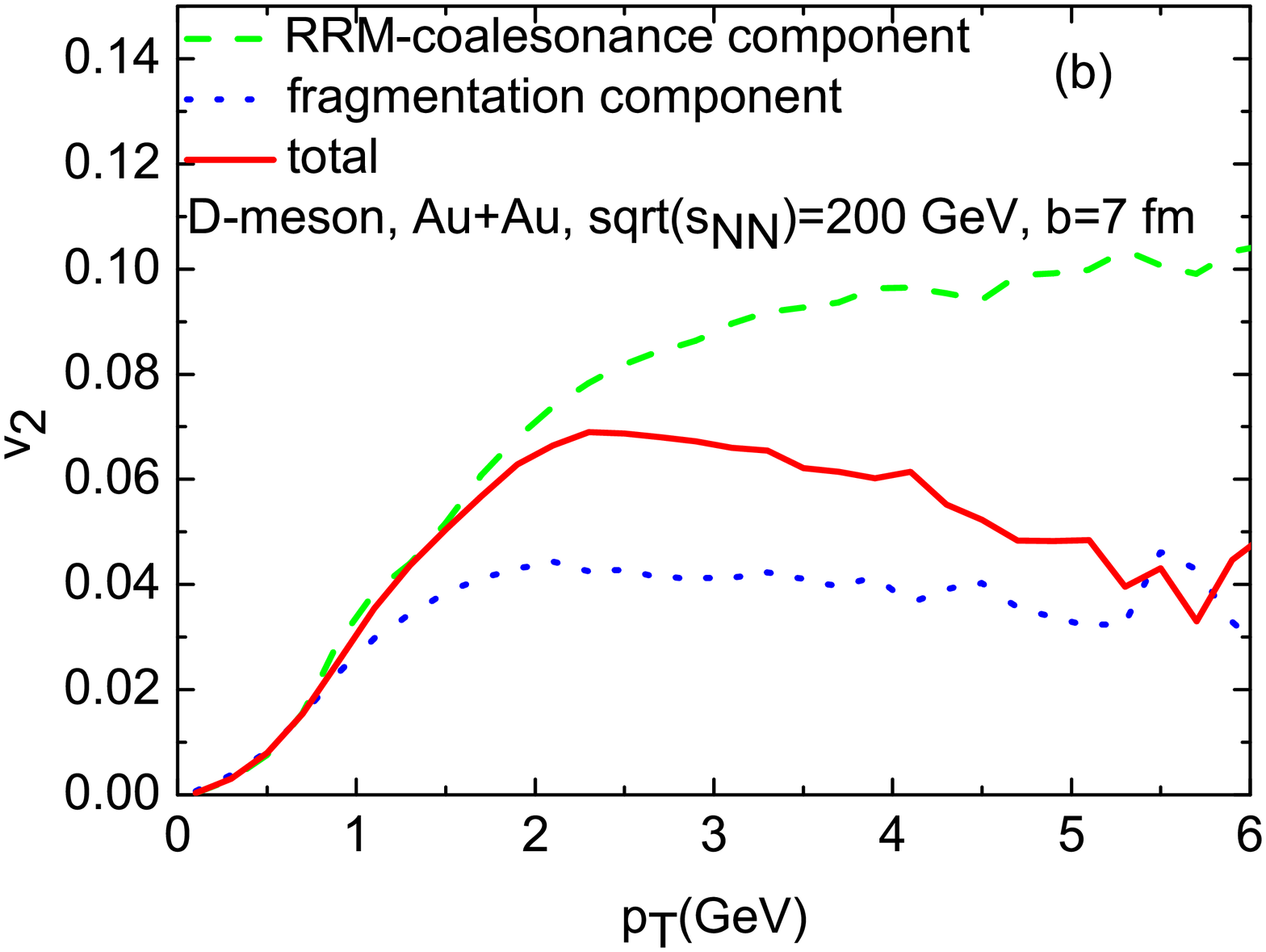} \caption{(Color online) (a) The coalescence,
fragmentation, and total $D$-meson $p_T$-spectrum for semi-central
($b$=7\,fm) Au+Au collisions at $\sqrt{s_{\rm NN}}=200~{\rm GeV}$,
normalized to one test particle. (b) The coalescence, fragmentation,
and total elliptic flow of $D$-mesons from the same calculation. }
\label{fig_D}
\end{figure}
\begin{figure}[!t]
\hspace{4mm}
\includegraphics[width=\columnwidth
]{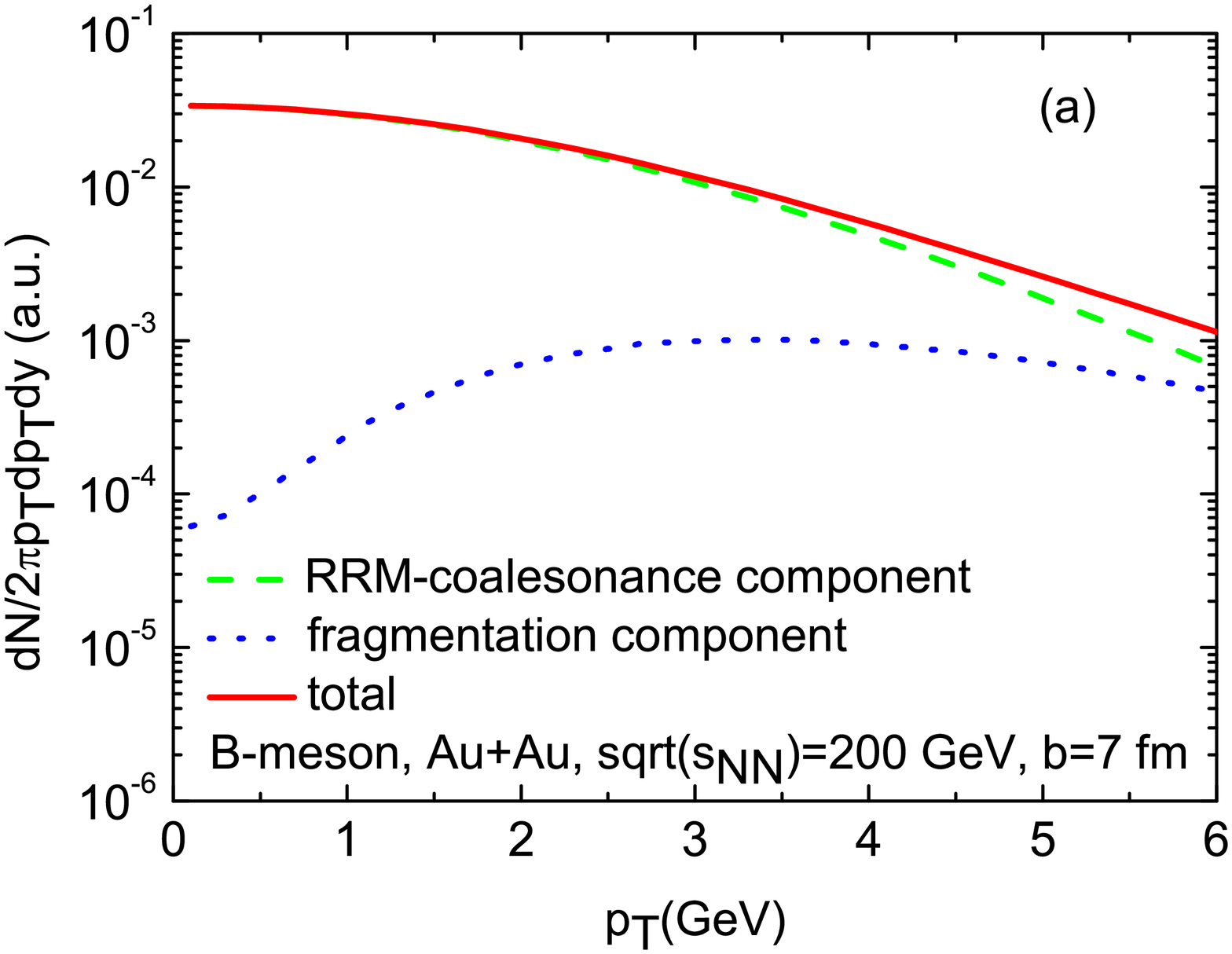}
\includegraphics[width=\columnwidth
]{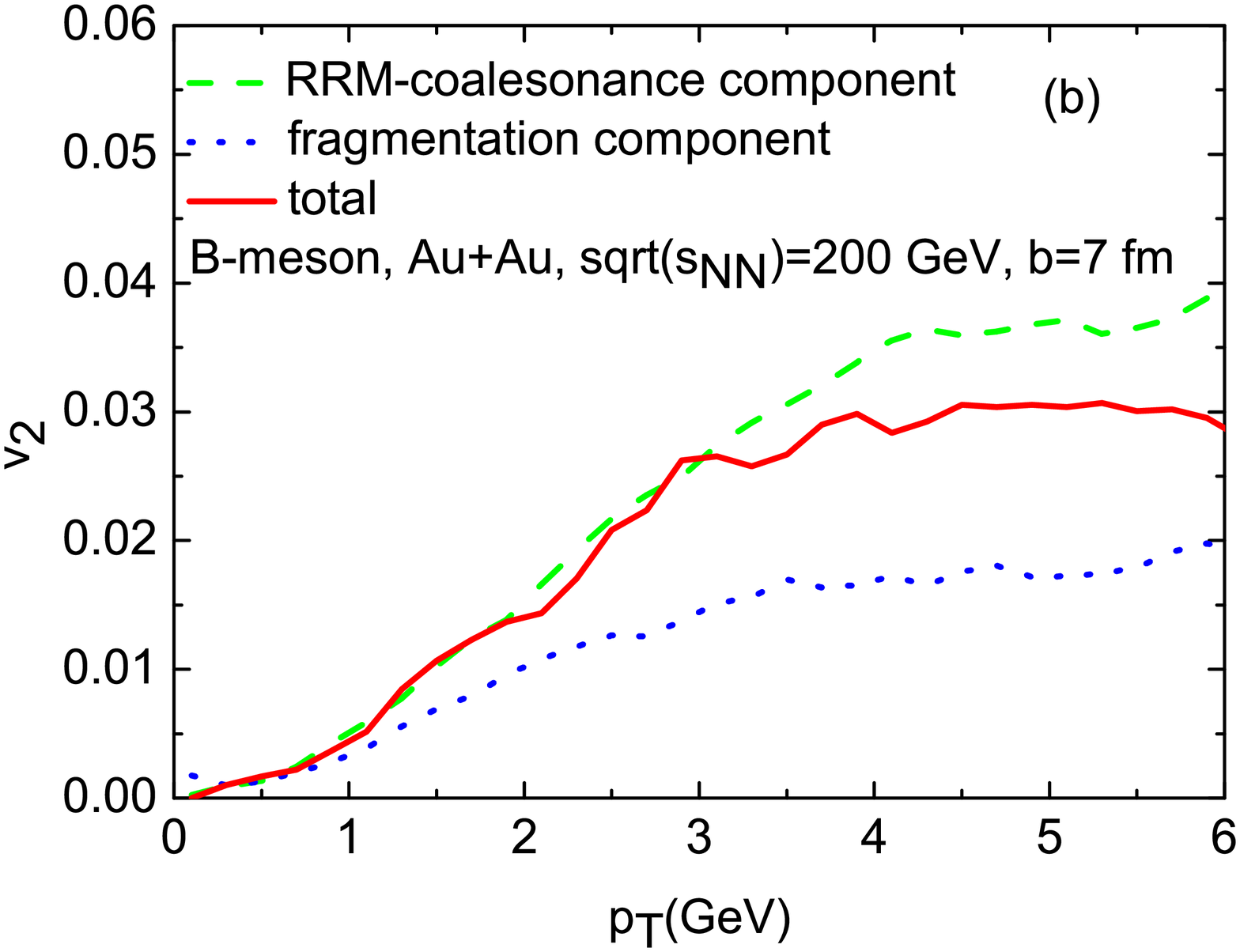} \caption{(Color online) The same as
Fig.~\ref{fig_D} but for $B$-mesons.}
\label{fig_B}
\end{figure}

Our final results for the $D$- and $B$-meson $p_T$-spectra and $v_2$
in semi-central Au+Au collisions at RHIC in the AZHYDRO+Langevin+RRM
approach are displayed in Figs.~\ref{fig_D} and \ref{fig_B},
respectively. The absolute norm of the spectra is arbitrary (we have
divided by the number of test particles, $5 \times 10^7$, used in
our simulations). We also show the individual recombination and
fragmentation components in the sum of Eq.~(\ref{eq:hfsum}). The
coalescence component dominates over fragmentation up to
$p_T\simeq3(7)$\,GeV for $D$ ($B$) mesons. Below these values,
coalescence with light quarks from the hydrodynamic heat bath
increases the $v_2$ by up to 30-40\% compared to the HQ $v_2$
(represented by the fragmentation component), whereas at high $p_T$
the heavy-meson spectra and $v_2$ approach the fragmentation values.

\begin{figure}[!t]
\hspace{4mm}
\includegraphics[width=\columnwidth
]{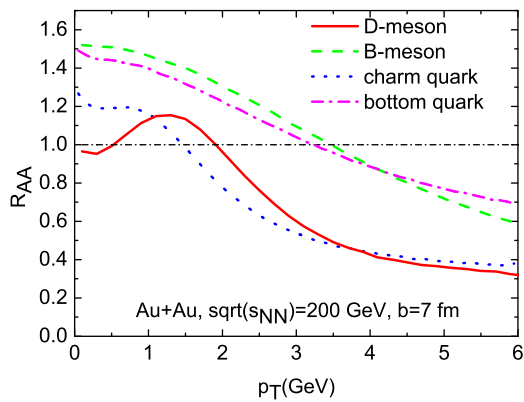}\caption{(Color online) $D$- and $B$-meson
nuclear modification factors for semi-central ($b$=7\,fm) Au+Au
collisions at RHIC. For comparison, the charm and bottom quark
$R_{\rm AA}$ are also shown.} \label{fig_RAA-DBc}
\end{figure}

Rescattering effects due to HQ diffusion in the QGP are also clearly
exhibited by the nuclear modification factor which we display in
Fig.~\ref{fig_RAA-DBc} for $D$- and $B$-mesons (as well as for
$c$ and $b$ quarks) in semi-central Au+Au collisions at RHIC. We note
a dip of the $D$-meson $R_{\rm AA}$ at low $p_T$, which is a consequence
of the standard mass effect in the collective flow of the $D$-mesons.
Remarkably, the dip is not present in the charm-quark $R_{AA}$,
underlining the effect of the larger $D$-meson mass and, more
importantly, the extra momentum added through coalescence with light
quarks. In a sense, coalescence acts as an additional interaction
driving the $D$-meson spectrum closer to equilibrium. The same effect is
most likely responsible for the slight suppression of the $D$-meson
spectrum below the $c$-quark spectrum at high $p_T$. Again, coalescence
acts as an additional interaction towards equilibration, which
in this case (high $p_T$) leads to suppression relative to the
$c$-quark spectrum.
This is, in fact, contrary to the naive belief that coalescence should
always add momentum to the formed-hadron spectrum. However, the approach
toward equilibrium is dictated by the underlying coalescence model
(here RRM) possessing the correct equilibrium limit.

The low-$p_T$ $D$-meson spectrum is rather sensitive to the
collective flow in the system. This is not the case for the
$B$-meson spectrum as it does not come close enough to equilibrium,
since the thermal $b$-quark relaxation times are relatively large
compared to the system's lifetime. In the following section we
quantify how varying the collective flow impacts charm ($D$-mesons)
spectra.

\section{Medium Flow Effect on $\mathbf D$-Meson Spectra}
\label{sec_flow}
In this section, we scrutinize the medium-flow effect on $D$-meson
spectra by comparing our results from the AZHYDRO background
with those from a more schematic fireball model.
We recall that AZHYDRO employs an EoS with a mixed phase of
appreciable time duration over which the sound velocity and thus the
acceleration vanish. This is incompatible with state-of-the-art
lattice QCD computations~\cite{Cheng:2007jq}. Consequently, AZHYDRO
presumably underestimates the flow at the end of the mixed phase and
around chemical freeze-out. Rather, it has been tuned to fit bulk
observables at kinetic freeze-out, $T_{\rm fo}$=100\,MeV, including
multi-strange baryons such as the $\Omega^-$~\cite{Kolb:2003dz}.
This is in conflict with the general belief (and empirical evidence)
that multi-strange hadrons ($\Omega$, $\Xi$, $\phi$) decouple close
to $T_c$ (due to their small hadronic rescattering cross sections)
but with rather large radial
flow~\cite{Adams:2005dq,He:2010vw,Mohanty:2009tz}, thus
corroborating our assertion of insufficient flow in AZHYDRO close to
$T_c$.

\begin{figure}[!t]
\hspace{4mm}
\includegraphics[width=\columnwidth
]{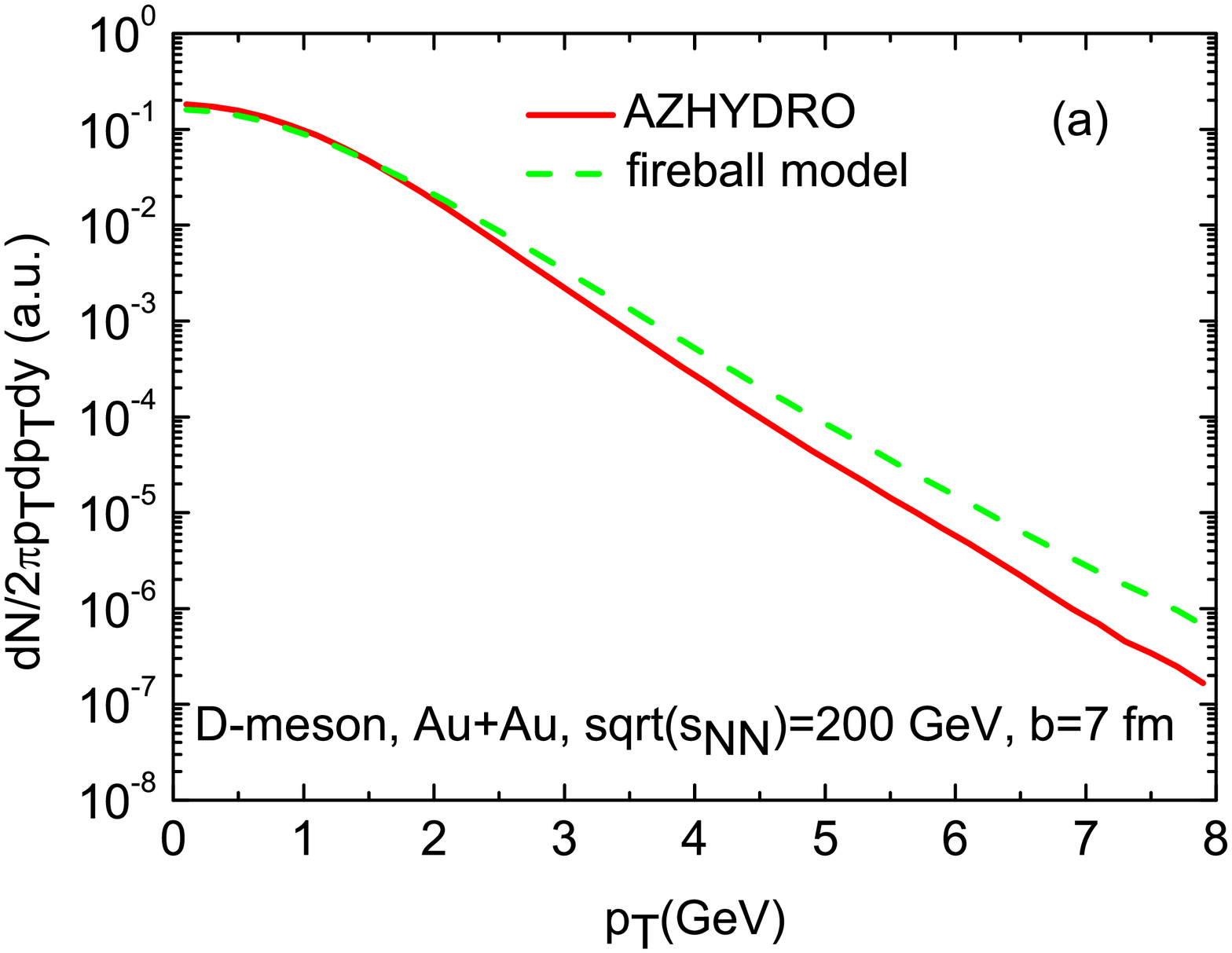}
\includegraphics[width=\columnwidth
]{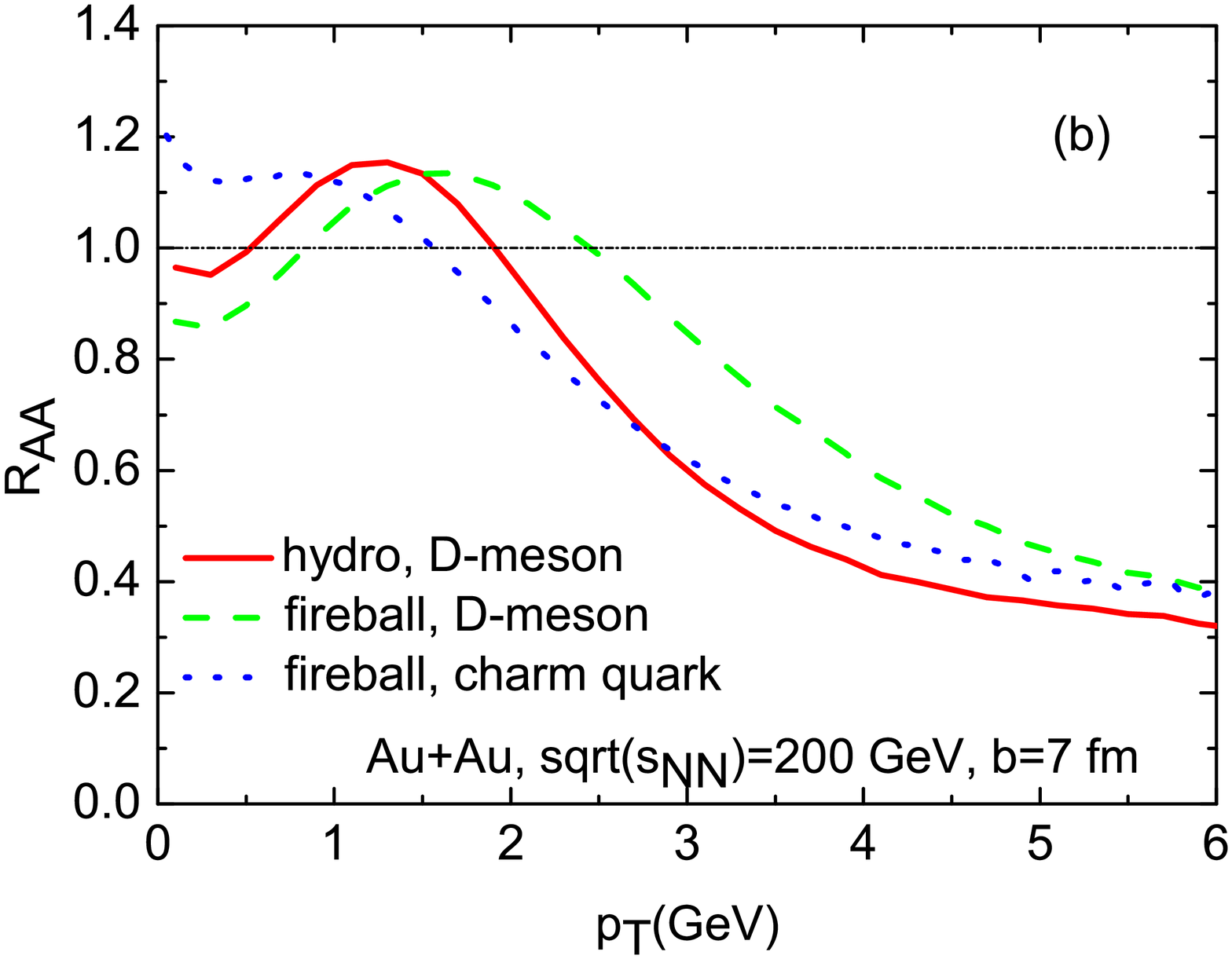}
\includegraphics[width=\columnwidth
]{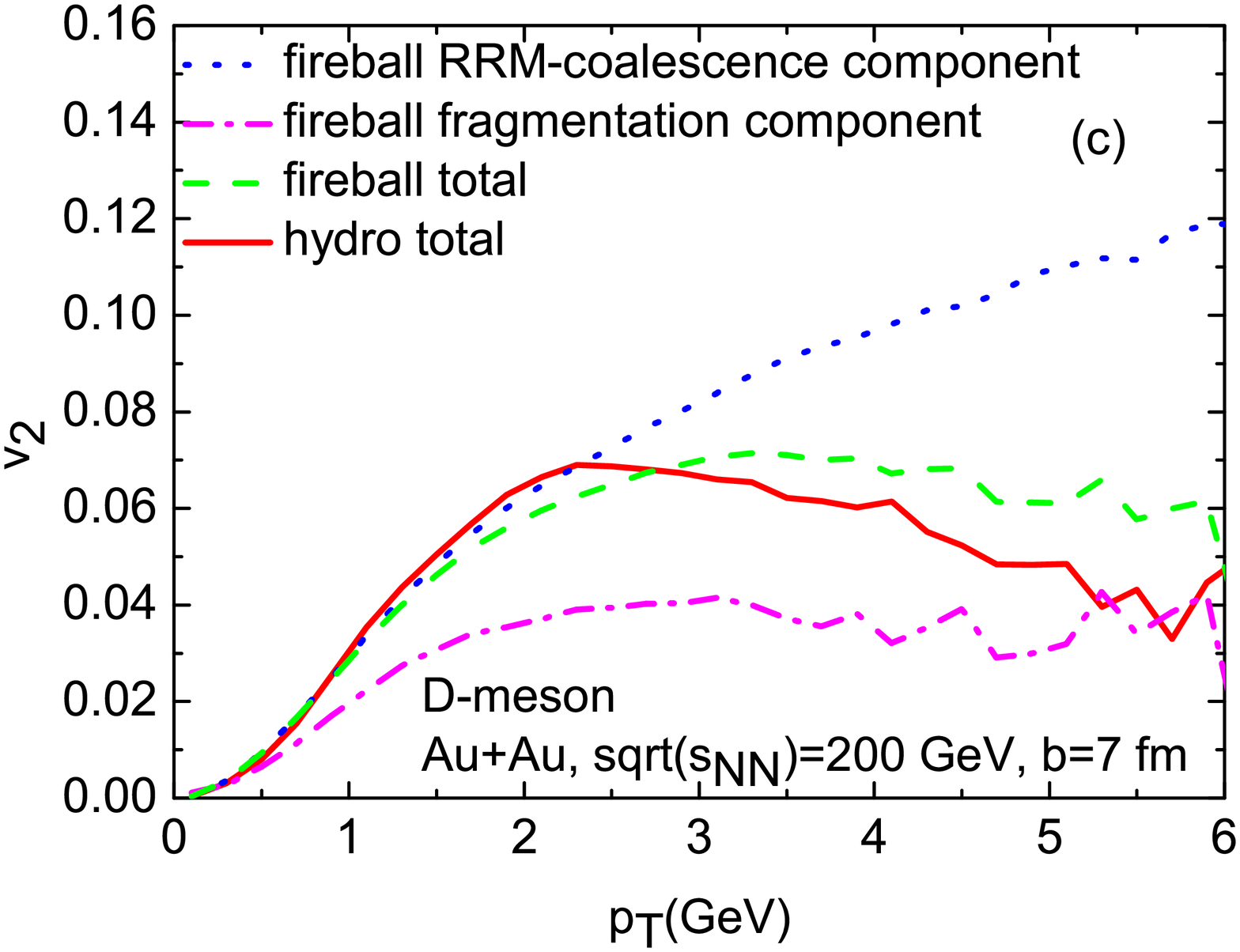} \caption{(Color online) Comparison of
$D$-meson spectra and $v_2$ from AZHYDRO and an elliptic fireball
with larger flow in semi-central ($b$=7\,fm) Au+Au collisions at
$\sqrt{s_{\rm NN}}$=200\,GeV.
(a) $p_T$-spectra from resonance recombination only (normalized to the
same total yield). (b) Nuclear modification factor for $D$-mesons
(coalescence + fragmentation) and charm quarks obtained with the
fireball model.
(c) Elliptic flow from coalescence, fragmentation and their weighted
sum for the fireball, and the total $v_2$ from AZHYDRO.}
\label{fig_hq-fb}
\end{figure}
Instead of retuning AZHYDRO with an improved EoS and/or initial
flow~\cite{Kolb:2002ve},
we here adopt a modest attitude by investigating flow effects on
$D$-meson spectra with a parameterized elliptic fireball model. We
have modified the fireball model introduced in
Ref.~\cite{vanHees:2005wb} such that the light-quark $p_t$-spectrum
and its integrated elliptic flow calculated at the end of the medium
evolution (QGP + mixed phase) agree with the empirical extraction of
Ref.~\cite{He:2010vw} where RRM was applied to experimental
multi-strange hadron spectra and $v_2$.
For consistency with the hydro framework we adopt Cooper-Frye
freeze-out rather than the Milekhin-like freeze-out in
Ref.~\cite{vanHees:2005wb} (cf.~Ref.~\cite{Gossiaux:2011ea}). The
retuned fireball results are included in Fig.~\ref{fig_AZHYDRO} in
direct comparison to those calculated in AZHYDRO. The integrated
bulk $v_2$ extracted from multi-strange particles ($4.99\%$) is
close to that calculated in AZHYDRO ($5.03\%$) at $T_c$, whereas the
extracted light-quark spectrum is much harder than in AZHYDRO, due
to the larger flow.

We have utilized the fireball evolution to perform charm-quark
Langevin simulations with our ``realistic" coefficients and to
compute $D$-meson spectra from RRM, summarized in
Fig.~\ref{fig_hq-fb}. As expected, the $D$-meson $p_T$-spectra,
shown in the upper panel for a hadronization with only
recombination, are significantly harder compared to using the
AZHYDRO background medium. The nuclear modification factor (middle
panel of Fig.~\ref{fig_hq-fb}) exhibits a more pronounced flow
effect at low momenta. The ``flow-bump" for heavy particles is
shifted to larger momenta in the fireball compared to AZHYDRO, and
the depletion toward $p_T=0$ is larger. A significant part of this
effect originates again from the coalescence process, as can be
inferred from comparing the change from the $c$-quark to $D$-meson
$R_{\rm AA}$ in the ``hard'' fireball model (middle panel of
Fig.\ref{fig_hq-fb}), relative to the ``softer'' AZHYDRO calculation
(Fig.~\ref{fig_RAA-DBc}), in the intermediate $p_T$-region. This is,
of course, due to the harder light-quark spectrum participating in
heavy-light recombination. Also note again that the coalescence
$D$-meson spectrum drops below the $c$-quark spectrum at high $p_T$,
which once more illustrates the role of resonance recombination as
an additional interaction driving the $D$-meson spectrum toward
equilibrium. The total $D$-meson $v_2$, shown in the lower panel of
Fig.\ref{fig_hq-fb}, receives a modest increase for
$p_T\gsim3$\,GeV, due to a larger $c$-quark coalescence probability
for stronger flow (comoving light partons have a higher phase-space
density at larger momentum when the flow is larger).

\section{Heavy Meson Semi-leptonic Decay and Observables}
\label{sec_electron}
\begin{figure}[!tb]
\hspace{4mm}
\includegraphics[width=\columnwidth
]{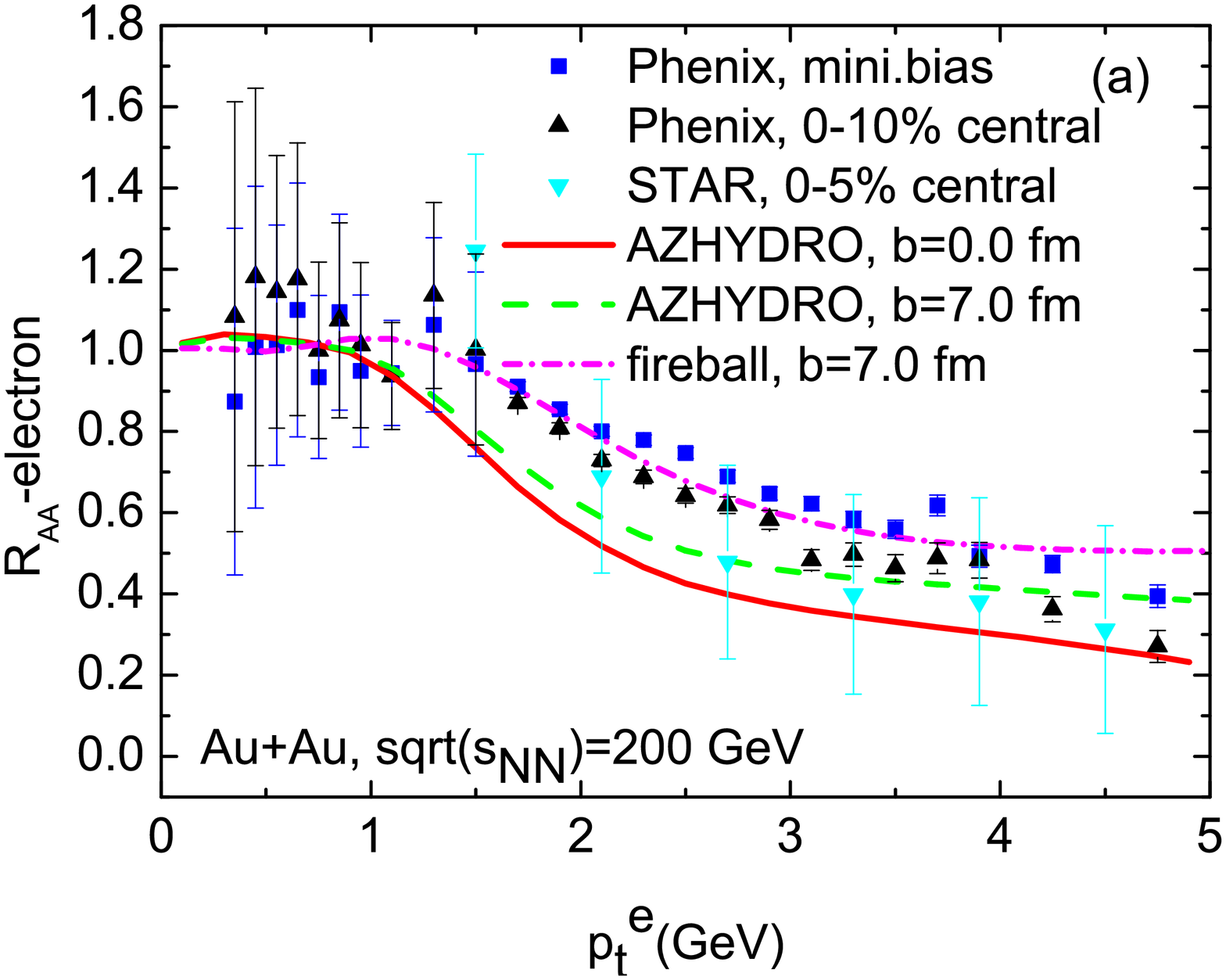}
\includegraphics[width=\columnwidth
]{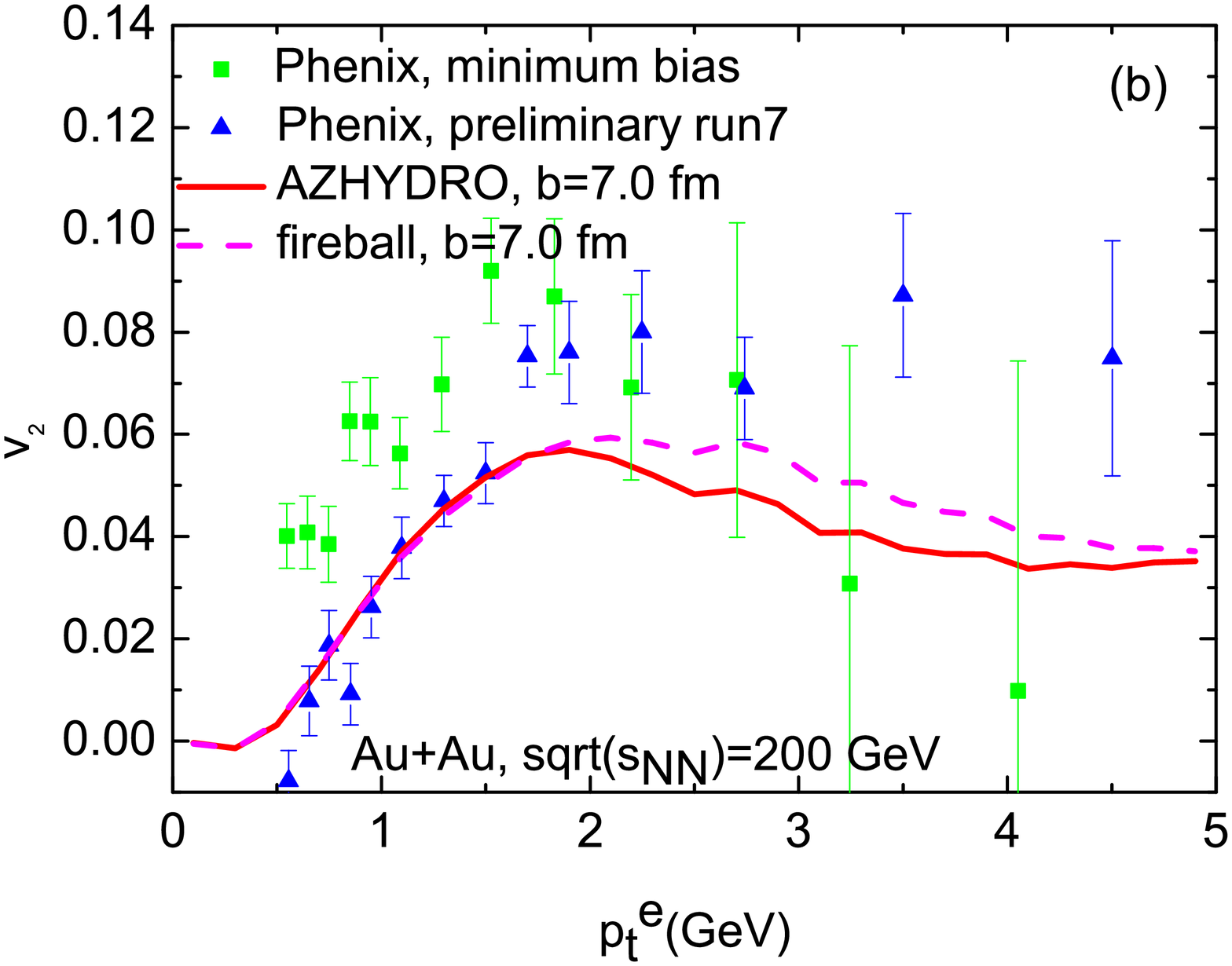}\caption{(Color online) (a) Electron $R_{\rm AA}$
produced from semi-leptonic $D$- and $B$-meson decays for central
and semi-central Au+Au collisions using hydro and fireball
background media, together with data from PHENIX for central and
minimum bias collisions~\cite{Adare:2006nq} and from STAR for
central collisions~\cite{Abelev:2006db}. (b) The same for electron
$v_2$. } \label{fig_e}
\end{figure}
Thus far at RHIC measurements of open HF in Au-Au collisions mostly
pertain to their semi-leptonic single-electron
decays~\cite{Abelev:2006db,Adare:2006nq,Adare:2010de}.\footnote{Only
very recently have direct $D$-meson data been reported.} The latter
have been shown to preserve the information on the nuclear
modification factor and elliptic flow of their parent hadrons rather
well~\cite{Greco:2003vf,Dong:2004ve}. In the following we treat
semi-leptonic decays of $D$- and $B$-mesons as free quark decays,
$c(b)\rightarrow s(c)+e+\nu_e$, albeit with effective quark masses
equal to their mesonic bound states to correctly account for phase
space: $m_b$=5.28\,GeV, $m_c$=1.87\,GeV, $m_s$=0.5\,GeV,
$m_e$=0.0005\,GeV and $m_{\nu}$=0 (light quarks are treated as
spectators).  We assume an average inclusive electronic branching
ratio of 11.5\% and 10.4\% for $c$ and $b$, respectively. We have
verified that hadronic form factors have little effect on the
electron energy spectrum in the parent-particle rest frame, as
already discussed in Ref.~\cite{Richman:1995wm}. Specifically, the
decays are performed via a Monte-Carlo simulation in the HQ rest
frame, with the 3-body phase space weighted by the decay matrix
element calculated in low-energy $V-A$
theory~\cite{Richman:1995wm,Griffiths1987}:
$\langle|\mathcal{M}|^2\rangle\propto (p_s\cdot p_{\nu})(p_c\cdot
p_e)$ for charm quarks, and $\langle|\mathcal{M}|^2\rangle\propto
(p_c\cdot p_e)(p_b\cdot p_{\nu})$ for bottom quarks. Subsequently,
the electron momentum is boosted to the lab frame using the
calculated heavy-meson spectrum.

Our numerical results for the single-electron $R_{\rm AA}(p_t^e)$
and $v_2(p_t^e)$ in Au+Au at RHIC, based on the $D$- and $B$-spectra
from our hydro+Langevin+RRM (as well as fireball+Langevin+RRM)
calculations, are compared to data~\cite{Abelev:2006db,Adare:2006nq}
in Fig.~\ref{fig_e}. For simplicity, we adopted an impact parameter
of $b$=7(0)\,fm to mimic the experimental minimum-bias (central)
results. On the one hand, the minimum bias sample for bulk
observables is closer to $b$=8-8.5\,fm, but, on the other hand, HQ
production scales more strongly with centrality, $\sim$A$^{4/3}$,
implying a smaller impact parameter for the HF minimum bias sample.
Our ``conservative" choice of somewhat smaller $b$ could thus
slightly overestimate the suppression and underestimate the $v_2$.

In the upper panel of Fig.~\ref{fig_e} one sees that the hydro-based
calculations overestimate the suppression found in the experimental
data for $R_{AA}^e$ in the regime where the uncertainty of the
latter is relatively small. The fireball calculations with larger
(and probably more realistic) flow improve on this aspect, which
reiterates the importance of medium collectivity in the
recombination process. For the elliptic flow (lower panel of
Fig.~\ref{fig_e}) we find good agreement of both calculations with
PHENIX run-7 data up to $p_t^e\simeq1.5$\,GeV, whereas the run-4
data are underpredicted. At larger $p_t^e$  our calculations tend to
underestimate the run-7 $v_2$ data. This indicates that additional
contributions to charm and bottom interactions are required, e.g.,
non-perturbative HQ-gluon scattering (presently we treat this part
perturbatively), radiative scattering and diffusion of HF hadrons in
the hadronic phase~\cite{Rapp:2009my}. We have recently estimated
the charm diffusion coefficient in the dense hadronic phase to be
comparable to the values in QGP that we use here~\cite{He:2011yi}.

\section{Summary and Conclusion}
\label{summary}
In this work, we have developed a framework in which heavy-quark
diffusion and hadronization in a quark gluon plasma are evaluated
consistently in a strongly coupled (non-perturbative) scenario. The
strong coupling is realized by resonance correlations which build up
in the hadronic channels of anti-/quark correlation functions as the
system cools down toward $T_c$. On the one hand, HQ transport has
been based on heavy-light quark $T$-matrices (consistent with vacuum
spectroscopy and the pQCD limit of high-energy scattering), in which
meson and diquark resonances enhance the HQ relaxation rate over
perturbative calculations~\cite{Riek:2010fk,Riek:2010py}; these
coefficients have been utilized in relativistic Langevin simulations
of HQ diffusion with a medium evolution described by an ideal
hydrodynamic model (which itself is based on the strong-coupling
limit). On the other hand, as the medium temperature drops towards
the critical value, the resonance correlations strengthen and
naturally trigger heavy-light quark recombination, which is carried
out in the RRM formalism. We have verified the equilibrium mapping
between quark- and meson-distributions on non-trivial hadronization
hypersurfaces in the hydrodynamic medium, which, in particular,
allowed us to identify and quantify the important effect of the
medium flow on the $D$-meson spectrum. We have also given a more
rigorous definition of the coalescence probability in terms of the
underlying formation rate, and in this way determined the partition
between recombination and fragmentation contributions to the
heavy-light spectra in absolute terms. We recover the mandatory
limits of equilibrium and independent fragmentation for low and
large transverse momenta, respectively.

Despite a few missing components in our HQ-transport and
hadronization scheme (as discussed below), we have carried our
calculations to the level of HF observabels in heavy-ion collisions.
In particular, we predict that the degree of charm-quark
thermalization is large enough to develop a characteristic
``flow-bump" in the $D$-meson nuclear modification factor. The
proper equilibrium limit of the underlying coalescence mechanism is
essential in developing this feature, while its location in $p_T$ is
sensitive to the strength of the medium flow. For $b$-quarks the
coupling to the medium is not strong enough for this feature to
emerge. We have further decayed our $D$- and $B$-meson spectra
semi-leptonically; the corresponding single-electron spectra show an
encouraging agreement with current RHIC data for $p_t^e\le 2$\,GeV;
the nuclear modification factor reiterates the importance of a
realistic flow strength, while the elliptic flow is a more sensitive
gauge for the magnitude of the HQ transport coefficient. Our current
estimate confirms earlier results for the spatial diffusion constant
in the vicinity of 5/(2$\pi T$).

Several uncertainties and areas of improvement remain, e.g., (i) a
more complete evaluation of HQ relaxation rates by including
non-perturbative effects in elastic scattering off gluons, as well
as adding radiative processes (expected to become relevant at high
$p_t$); (ii) hadronic diffusion, especially in light of recent
estimates indicating comparable strength of HF transport in the
hadronic and QGP phases near $T_c$~\cite{He:2011yi}; (iii) a more
complete hadro-chemistry at $T_c$ including strange $D$-mesons and
charmed baryons; (iv) a possible Cronin effect in the initial HQ
spectra, which mostly affects $R_{AA}$; (v) further improvements in
the medium evolution, i.e., a hydrodynamic background with harder
spectra (e.g., due to viscosity, initial flow or a harder EoS). The
framework put forth in this paper allows us to systematically
address all of these items. This will not only be beneficial for the
open heavy-flavor sector (and an ultimate determination of the QGP's
viscosity) but also for its impact on closely related observables
such as heavy quarkonia and intermediate-mass dileptons.

\vspace{0.3cm}

\acknowledgments We gratefully acknowledge helpful discussions with
H.~van Hees, V.~Greco and X.~Zhao, and we would like to thank E.~Frodermann
for his support with AZHYDRO. We are indebted to F.~Riek for providing
his results for the HQ transport coefficients. This work was supported
by the U.S.\ National Science Foundation (NSF) through CAREER grant PHY-0847538
and grant PHY-0969394, by the A.-v.-Humboldt
Foundation, by the RIKEN/BNL Research Center and DOE grant
DE-AC02-98CH10886, and by the JET Collaboration and
DOE grant DE-FG02-10ER41682.

\end{document}